\documentclass[useAMS,usenatbib]{mn2e} 

\usepackage{mathptmx}
\usepackage[british]{babel}
\usepackage{amsmath,amssymb}
\usepackage[pdftex]{graphicx}
\usepackage{microtype}
\title[Relativistic effects on tidal disruption kicks]{\bf Relativistic effects on tidal disruption kicks of solitary stars}
\author[E. Gafton, E. Tejeda, J. Guillochon, O. Korobkin, S. Rosswog]{E. Gafton$^{1}$\thanks{E-mail:
emanuel.gafton@astro.su.se}, E. Tejeda$^{1}$, J. Guillochon$^{2,3}$, O. Korobkin$^{1}$ and S. Rosswog$^{1}$\\
$^{1}$ The Oskar Klein Centre, Department of Astronomy, 
AlbaNova, Stockholm University, SE-106 91 Stockholm, Sweden\\
$^{2}$ Harvard-Smithsonian Center for Astrophysics, Institute for Theory and Computation, 60 Garden Street, Cambridge, MA 02138, USA\\
$^{3}$ Einstein Fellow}

\newcommand{\rt}{r_\s{\mathrm{s}}}

\def\be{\begin{equation}}
\def\ee{\end{equation}}
\def\bi{\begin{itemize}}

\def\ei{\end{itemize}}
\def\ben{\begin{enumerate}}
\def\een{\end{enumerate}}
\def\bea{\begin{eqnarray}}
\def\eea{\end{eqnarray}}

\def\edo{\end{document}}

\newcommand{\cc}{\mathrm{c}}

\newcommand{\G}{\mathrm{G}}
\newcommand{\s}[1]{{\text{\scriptsize $#1$ }}\hspace{-2pt}}
\newcommand{\rs}{r_\s{\mathrm{s}}}

\renewcommand{\rt}{r_\s{\mathrm{t}}}
\newcommand{\rp}{r_\s{\mathrm{p}}}

\newcommand{\eq}[1]{Eq.\,\eqref{#1}}
\newcommand{\eqp}[1]{(Eq.\,\ref{#1})}
\newcommand{\eqs}[1]{Eqs.\,\eqref{#1}}
\newcommand{\MGRO}{MGRO} 

\defcitealias{manukian13}{MGRO}
\defcitealias{tejeda13}{TR}

\begin{document}

\date{Accepted . Received ; in original form 2014 October 17} 
\pagerange{\pageref{firstpage}--\pageref{lastpage}} \pubyear{2015} 

\maketitle

\label{firstpage} 

\begin{abstract}
Solitary stars that wander too close to their galactic centres can become
tidally disrupted, if the tidal forces due to the supermassive black hole (SMBH)
residing there overcome the self-gravity of the star.
If the star is only partially disrupted, so that a fraction survives
as a self-bound object, this remaining core will experience a net gain
in specific orbital energy, which translates into a velocity ``kick'' of up
to $\sim 10^3$ km/s.

In this paper, we present the result of smoothed particle
hydrodynamics (SPH) simulations of such partial disruptions, 
and analyse the velocity kick imparted on the surviving core. We compare 
$\gamma=5/3$ and $\gamma=4/3$ polytropes disrupted in both a Newtonian potential, and 
a generalized potential that reproduces most relativistic effects around a Schwarzschild
black hole either exactly or to excellent precision. 
For the Newtonian case, we confirm the results of previous studies 
that the kick velocity of the surviving core is 
virtually independent of the ratio of the black hole to stellar mass, and is a function of the 
impact parameter $\beta$ alone, reaching at most the escape velocity of the original 
star.
For a given $\beta$, relativistic effects become increasingly important for larger black hole masses. 
In particular, we find that the kick velocity increases with the black hole mass, making larger kicks more common than in the Newtonian case, as low-$\beta$ encounters are statistically more likely than high-$\beta$ encounters.

The analysis of the tidal tensor for the generalized potential shows that our results
are robust lower limits on the true relativistic kick velocities, and are generally in very good agreement 
with the exact results.
\end{abstract}


%
%
\section{Introduction}\label{sec:introduction}
The centres of most galaxies are hosts to supermassive black holes (SMBHs)
with masses ranging from $\sim 10^5~M_\odot$ \citep{secrest12}
to as much as a few $\times 10^{10}~M_\odot$ \citep{vandenbosch12}.
The SMBH's mass is comparable to the combined mass of all the stars in the nucleus, 
and together they control the orbital dynamics of individual stars. Various mechanisms 
(the most important being two-body scattering; see e.g. \citealt{alexander05}) 
may at times bring one of the stars onto a nearly radial orbit that reaches the immediate
vicinity of the SMBH, with the periapsis distance $\rp$ becoming comparable
to the tidal radius $\rt\equiv (M_{\rm bh}/m_\star)^{1/3} r_\star$ \citep{frank78}.
On average, this happens at a rate of $\sim 10^{-5}$ yr$^{-1}$ per galaxy  
\citep{magorrian99,wang04}. 
The strength of the encounter, quantified by the impact parameter $\beta\equiv \rt/\rp$,
will ultimately determine how much mass the star loses due to tidal interactions, 
whether it survives ($\beta\lesssim 1$) or is completely ripped apart ($\beta\gtrsim 1$)
\citep{rees88}.

Recent simulations by \citet[][henceforth MGRO; see \citealt{manukian14} for errata]{manukian13} showed that in partial disruptions,
in which a fraction of the star survives as a self-bound object, the remaining core 
may receive a boost in specific orbital energy, corresponding
to a velocity ``kick'' that can reach the surface escape velocity of the original star, 
and may later on be observed as a ``turbo-velocity star''.
The source of this increase in energy has been linked to the asymmetry of the
two tidal tails created during the disruption process, which by conservation of linear
momentum boosts the velocity of the surviving core. 
This asymmetry (and the velocity kick it induces) 
appears to be an inherent property of tidal disruption events, 
and has also been observed in tidal disruptions 
of planets \citep{faber05}, white dwarfs \citep{cheng13} and neutron stars \citep{rosswog00,kyutoku13}.

In this paper we extend the previous results by performing smoothed
particle hydrodynamics (SPH) simulations of 
tidal disruptions of solar-type stars by SMBHs. We use both a completely Newtonian approach
and one where the orbital dynamics around a Schwarzschild black hole is accurately 
reproduced by a generalized potential \citep[][henceforth TR]{tejeda13}. Apart from a verification
of the previous results with different numerical methods, our main goal is to
quantify to which extent the relativistic effects from a Schwarzschild black hole
would impact on the final velocities.
The study of these events may contribute to the understanding of hyper-velocity stars 
(HVSs; \citealt{hills88}), which were thought to result either from the tidal disruption of binary stars,
or by scattering off the stellar-mass black holes segregated in the Galactic centre \citep{oleary08}.
While these processes can indeed impart velocities in excess of $10^3$ km/s 
\citep{brown05,antonini10}, most observed HVSs seem to have rather modest velocities,
typically around $\sim 400$ km/s \citep[e.g.][]{palladino14,zheng14,zhong14}. A number of HVSs have even
been observed on bound orbits around the Galactic centre, but with sufficiently high velocities so as to
constitute a distinct population of velocity outliers \citep{brown07}. 
Tidal disruptions of solitary stars may therefore yield sufficiently large kicks to explain many of these 
HVSs, especially if the kicks are significantly enhanced by relativistic effects.

\section{Method}\label{sec:method}
In our simulations we use the Newtonian SPH code described in detail by \citet{rosswog09},
with self-gravity computed using a binary tree similar to that of 
\citet{benz90}. The tree accuracy parameter (i.e., the opening angle $\theta> {H_B}/{R_{AB}}$ which controls whether a distant tree node $B$, of size $H_B$ and 
located at a distance $R_{AB}$ from node $A$, is allowed to act as a multipole source of gravity for node
$A$ or needs to be further resolved into its constituents) was set to $\theta=0.5$, 
corresponding to an average relative force error of $\lesssim 0.1\%$.
All simulations use $10^5$ SPH particles, unless otherwise stated
(we verified the results via two runs with $10^6$ particles).

\begin{table*}
\begin{minipage}{120mm}
 \caption{Overview of the SPH simulations discussed in this paper, grouped into three categories: 59 core simulations (1--59) with $\gamma=5/3$ and $10^5$ SPH particles, covering the entire range of $q$ and $\beta$ discussed in the paper; 25 test simulations (60--84) with $\gamma=5/3$ (of which 24 with $10^6$ SPH particles, and 1 with $\theta=0.2$); 21 test simulations (85--105) with $\gamma=4/3$. 
 For each simulation we show the polytropic index $\gamma$, the number of SPH particles $N_{\rm
   part}$, the ratio $q=M_{\rm bh}/m_\star$, the potential $\Phi$
 (which is either the Newtonian potential, $\Phi_{\rm N}$, or the
 generalized Newtonian (TR) potential, $\Phi_{\rm TR}$), and the impact
 parameter $\beta=\rt/\rp$.}
 \label{table:overview}
 \begin{center}
 \begin{tabular}{l|ccrlcl}
 \hline
  Number & $\gamma$ & $N_{\rm part}$ & $q$ & $\Phi$ & $\theta$ & $\beta$ \\
 \hline
   1--7    & 5/3 & $10^5$ &         $10^6$ & $\Phi_{\rm N}$ & 0.5 & 0.60, 0.65, 0.70, 0.75, 0.80, 0.85, 0.90 \\
   8--16   & 5/3 & $10^5$ &         $10^6$ & $\Phi_{\rm TR}$ & 0.5 & 0.60, 0.65, 0.70, 0.75, 0.80, 0.85, 0.88, 0.89, 0.90 \\
  17--23   & 5/3 & $10^5$ & $4\times 10^6$ & $\Phi_{\rm N}$ & 0.5 & 0.60, 0.65, 0.70, 0.75, 0.80, 0.85, 0.90 \\
  24--31   & 5/3 & $10^5$ & $4\times 10^6$ & $\Phi_{\rm TR}$ & 0.5 & 0.60, 0.65, 0.70, 0.75, 0.80, 0.83, 0.84, 0.85 \\
  32--38   & 5/3 & $10^5$ &         $10^7$ & $\Phi_{\rm N}$ & 0.5 & 0.60, 0.65, 0.70, 0.75, 0.80, 0.85, 0.90  \\
  39--45   & 5/3 & $10^5$ &         $10^7$ & $\Phi_{\rm TR}$ & 0.5 & 0.60, 0.65, 0.70, 0.75, 0.77, 0.78, 0.80 \\
  46--52   & 5/3 & $10^5$ & $4\times 10^7$ & $\Phi_{\rm N}$ & 0.5 & 0.60, 0.65, 0.70, 0.75, 0.80, 0.85, 0.90 \\
  53--59   & 5/3 & $10^5$ & $4\times 10^7$ & $\Phi_{\rm TR}$ & 0.5 & 0.60, 0.65, 0.68, 0.70, 0.71, 0.72, 0.75 \\
\hline
  60--66   & 5/3 & $10^6$ &         $10^6$ & $\Phi_{\rm N}$ & 0.5 & 0.60, 0.65, 0.70, 0.75, 0.80, 0.85, 0.90  \\
  67--73   & 5/3 & $10^6$ &         $10^6$ & $\Phi_{\rm TR}$ & 0.5 & 0.60, 0.65, 0.70, 0.75, 0.80, 0.85, 0.90  \\
  74--75   & 5/3 & $10^6$ & $4\times 10^6$ & $\Phi_{\rm TR}$ & 0.5 & 0.83, 0.84 \\
  76       & 5/3 & $10^6$ &         $10^7$ & $\Phi_{\rm N}$ & 0.5 & 0.70 \\
  77       & 5/3 & $10^6$ &         $10^7$ & $\Phi_{\rm TR}$ & 0.5 & 0.70 \\  
  78--83   & 5/3 & $10^6$ & $4\times 10^7$ & $\Phi_{\rm TR}$ & 0.5 & 0.60, 0.65, 0.67, 0.70, 0.71, 0.72 \\
  84       & 5/3 & $10^5$ &         $10^7$ & $\Phi_{\rm TR}$ & 0.2 & 0.70 \\
\hline
  85       & 4/3 & $10^6$ &         $10^4$ & $\Phi_{\rm TR}$ & 0.5 & 1.60 \\
  86       & 4/3 & $10^6$ &         $10^6$ & $\Phi_{\rm TR}$ & 0.5 & 1.30 \\
  87--91   & 4/3 & $10^5$ &         $10^6$ & $\Phi_{\rm TR}$ & 0.5 & 1.45, 1.55, 1.65, 1.75, 1.80 \\
  92--95   & 4/3 & $10^5$ &         $10^6$ & $\Phi_{\rm N}$ & 0.5 & 1.45, 1.55, 1.65, 1.75 \\
  96--100  & 4/3 & $10^5$ & $4\times 10^6$ & $\Phi_{\rm N}$ & 0.5 & 1.10, 1.20, 1.30, 1.70, 1.80 \\
 101--105  & 4/3 & $10^5$ & $4\times 10^6$ & $\Phi_{\rm TR}$ & 0.5 & 1.10, 1.20, 1.30, 1.65, 1.70 \\
 \hline
 \end{tabular}\end{center}
\end{minipage}
\end{table*}

We model stars as polytropic fluids with $\gamma=5/3$, which initially
satisfy the Lane-Emden equation for $m_\star=1~M_\odot$ and $r_\star=1~R_\odot$.
It has long been known that $\gamma=5/3$ polytropes are disrupted
at smaller $\beta$'s than $\gamma=4/3$ polytropes, which, being more centrally condensed,
are able to survive deeper encounters (see e.g. \citealt{guillochon13}).
In order to compare our results with those of \citetalias{manukian13}, who used $\gamma=4/3$, 
we also perform a few test simulations in which the initial 
stellar profiles are given by $\gamma=4/3$ polytropes, 
but the fluid, being gas-pressure dominated, reacts to dynamical
compressions and expansions according to a $\gamma=5/3$ equation of state.

The black hole gravity is modelled with both
the Newtonian potential (``$\Phi_{\rm N}$'') and the generalized
Newtonian potential (``$\Phi_{\rm TR}$''; \citealp{tejeda13}),
\begin{equation}
\Phi_{\rm TR}(r,\dot{r},\dot{\varphi})=
-\frac{GM_{\rm bh}}{r}-\frac{1}{2}\left(\frac{\rs}{r-\rs}\right)
\left[
\left(\frac{2r-\rs}{r-\rs}\right)
\dot{r}^2
+r^2\dot{\varphi}^2
\right],
\end{equation} where $\rs=2\,\G M_{\rm bh}/\cc^2$ is the Schwarzschild radius. The potential is based on an
expansion of the relativistic equations of motion in the low-energy limit, without necessarily
implying low-velocities or weak field.
It has been shown to reproduce essentially all relevant orbital properties 
around a Schwarzschild black hole either exactly or to a very high degree of accuracy.

In the \citetalias{tejeda13} potential, the specific relativistic orbital energy $\epsilon_{\rm TR}$
is computed as
\begin{equation}
\epsilon_{\rm TR}=
\frac{1}{2}\left[
\frac{r^2 \dot{r}^2}{(r-\rs)^2}+
\frac{r^3 \dot{\varphi}^2}{r-\rs}
\right]-\frac{GM_{\rm bh}}{r}.
\end{equation} The self-bound mass is calculated using
the iterative energy-based approach described by \citet[][Sec. 2.2]{guillochon13}, 
with the gravitational self-potential calculated using a fast binary tree \citep{gafton11}. 
The kick velocity of the self-bound object (at infinity) is then computed as 
\begin{equation}\label{eq:vkick}
v_{\rm kick}=\sqrt{2(\epsilon-\epsilon_0)},
\end{equation} where $\epsilon_0$ is the specific orbital energy of the initial star
at the beginning of the simulation, and $\epsilon$ is the specific orbital energy
of the self-bound remnant. This definition does not take into account the underlying
galactic gravitational potential, and therefore the kick velocity is expected to decrease as 
stars ``climb'' out of the galactic potential.

Since we are only considering parabolic orbits,
$\epsilon_0$ is approximately equal to zero, with small numerical deviations due to the fact that
only the centre of mass is on a truly parabolic orbit, since we impart the same 
initial orbital velocity to all particles, while starting with the star at an 
initial distance of $5~\rt$ from the black hole, instead of at infinity.

We simulate encounters with mass ratios $q\equiv M_{\rm bh}/m_\star$ 
in the range $10^{6} \leq q \leq 4\times 10^{7}$, and impact parameters $\beta$ in the range
$0.6 \leq \beta \leq 0.9$. 
For the $\gamma=4/3$ runs we use the same values of $\beta$ and $q$ as \citetalias{manukian13}, namely
1.0 to 1.8, and $10^3$ to $10^6$, respectively.
The values for $\beta$ are chosen so that a self-bound core always survives,
while $q$ is chosen so that the star is disrupted outside the Schwarzschild
radius of the SMBH.

A summary of all the simulations performed for this paper is presented
in Table~\ref{table:overview}.

\section{Results}\label{sec:results}
\subsection{Disruption dynamics}
\begin{figure*}
\centering
\includegraphics[width=\linewidth]{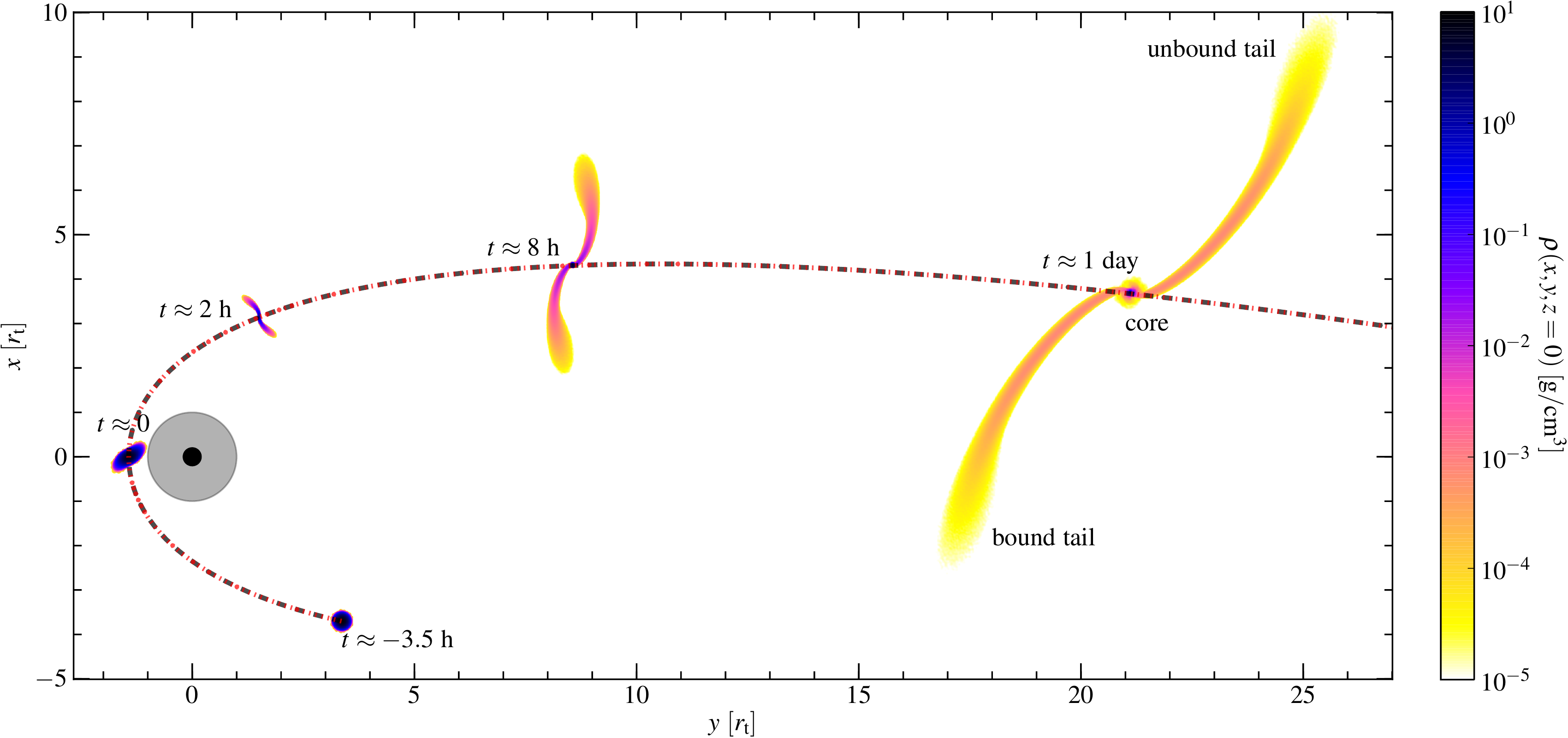}
\caption{Evolution of the stellar fluid during a typical, parabolic partial tidal disruption.
The plot shows a cross-section of the density profile in the orbital plane ($z=0$).
Here, $N_{\rm part}=10^6$, $q=10^7$, $\beta=0.7$, and $\Phi=\Phi_{\rm TR}$.
The Schwarzschild radius (black disc), tidal radius (gray disc), geodesic trajectory of the centre of mass
in Schwarzschild spacetime (dotted red line) and trajectory of the centre of mass obtained with the 
TR potential (dashed black line) are shown to scale.
Due to the spatial scales involved, we show the stellar debris magnified by a factor of 50 before periapsis passage ($t\leq 0$) and magnified by a factor of 10 afterwards ($t>0$).
Note that due to relativistic periapsis shift the orbit is not a parabola.
}
\label{fig:config}
\end{figure*}

The typical evolution of the stellar fluid during a partial tidal disruption is shown in
Fig.~\ref{fig:config}. As it approaches periapsis, the star is heavily spun-up and distorted,
being stretched in the radial direction, corresponding to the one positive eigenvalue
of the tidal tensor, and compressed in the azimuthal and vertical directions (see \citealt{luminet86}; 
see also Appendix~\ref{sec:app} for a discussion on the tidal tensor).
As the star overfills its Roche lobe it starts to shed mass through
the Lagrangian points L1 and L2, forming a bound and an unbound (to the SMBH) tail,
respectively.
As the star is receding from the black hole, the tails and the core stop exchanging energy and 
angular momentum (i.e., the energies become become ``frozen-in'')
and the core recollapses into a self-bound, spherical object.

\begin{figure*}
\centering
\includegraphics[width=\linewidth]{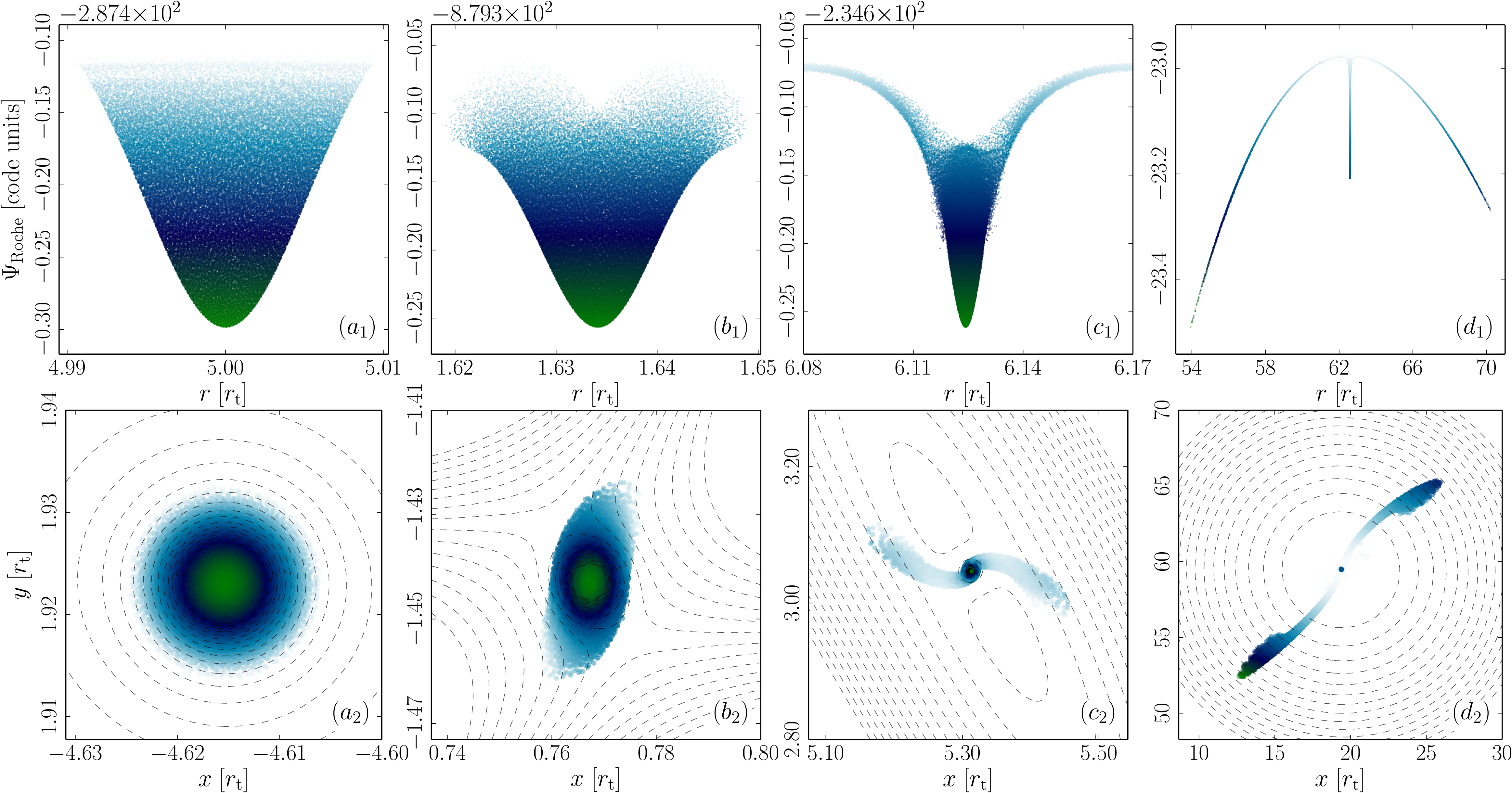}
\caption{A typical, time-varying Roche potential in a partial tidal disruption exhibits a number of stages, shown here
as snapshots at:
(a) $t\approx -3.5$~h before periapsis passage, i.e. at the beginning of the simulation, when the star is self-bound;
(b) $t\approx 20$~min, just as the first particles exit the Roche lobe of the star and start forming the tidal tails (bound tail first);
(c) $t\approx 4$~h, as the energy distribution of the debris starts to freeze and the core and tails approach their final masses;
(d) $t\approx 4$~days after the disruption.
This simulation used $N_{\rm part}=10^5$, $q=10^6$, $\beta=0.65$, and $\Phi=\Phi_{\rm N}$.
\textit{Upper row.} $\Psi_{\rm Roche}(r)$, where $r$ is the distance to the SMBH.
The values of $\Psi_{\rm Roche}$ are given in code units and -- to simplify axis labelling -- are offset 
by the values shown in the upper left corner of the panels.
\textit{Lower row.} Contours of $\Psi_{\rm Roche}(x,y)$ in the orbital plane, with the particles overplotted
and coloured according to the value of the potential. All coordinate
axes use a global Cartesian coordinate system, with the SMBH always located at
$(0,0,0)$, and the stellar fluid moving along a parabolic orbit around
it. In the lower row we are essentially ``zooming in'' on the SPH
particle distribution as it first approaches, and then recedes from the SMBH.
}
\label{fig:Roche}
\end{figure*}

Fig.~\ref{fig:Roche} shows the locations of the SPH particles in the 
time-varying Roche potential with respect to the centre of mass of the star 
($\Psi_{\rm Roche}$, see Appendix~\ref{sec:app2} for a derivation), 
calculated along the radial direction (upper row), 
and contours of $\Psi_{\rm Roche}$ in the orbital plane (lower row). 
The formation of the two tidal
tails is asymmetric from the beginning (the star first overflows its Roche lobe through L1), but it is
most clearly seen in panel (d$_2$).
It is interesting to observe that in panel (b$_2$), representative of the star during
the actual disruption, the Roche potential is not aligned with the star, i.e.,
the points through which the star sheds mass are not always aligned with the
instantaneous L1 and L2. This occurs because the orbital time scale of the system is shorter than the
dynamical time scale on which the fluid can react to the extremely fast-changing Roche 
potential ($\tau_{\rm orb} < \tau_{\rm dyn}$).

\subsection{Self-bound mass}
\begin{figure*}
\includegraphics[width=.48\linewidth]{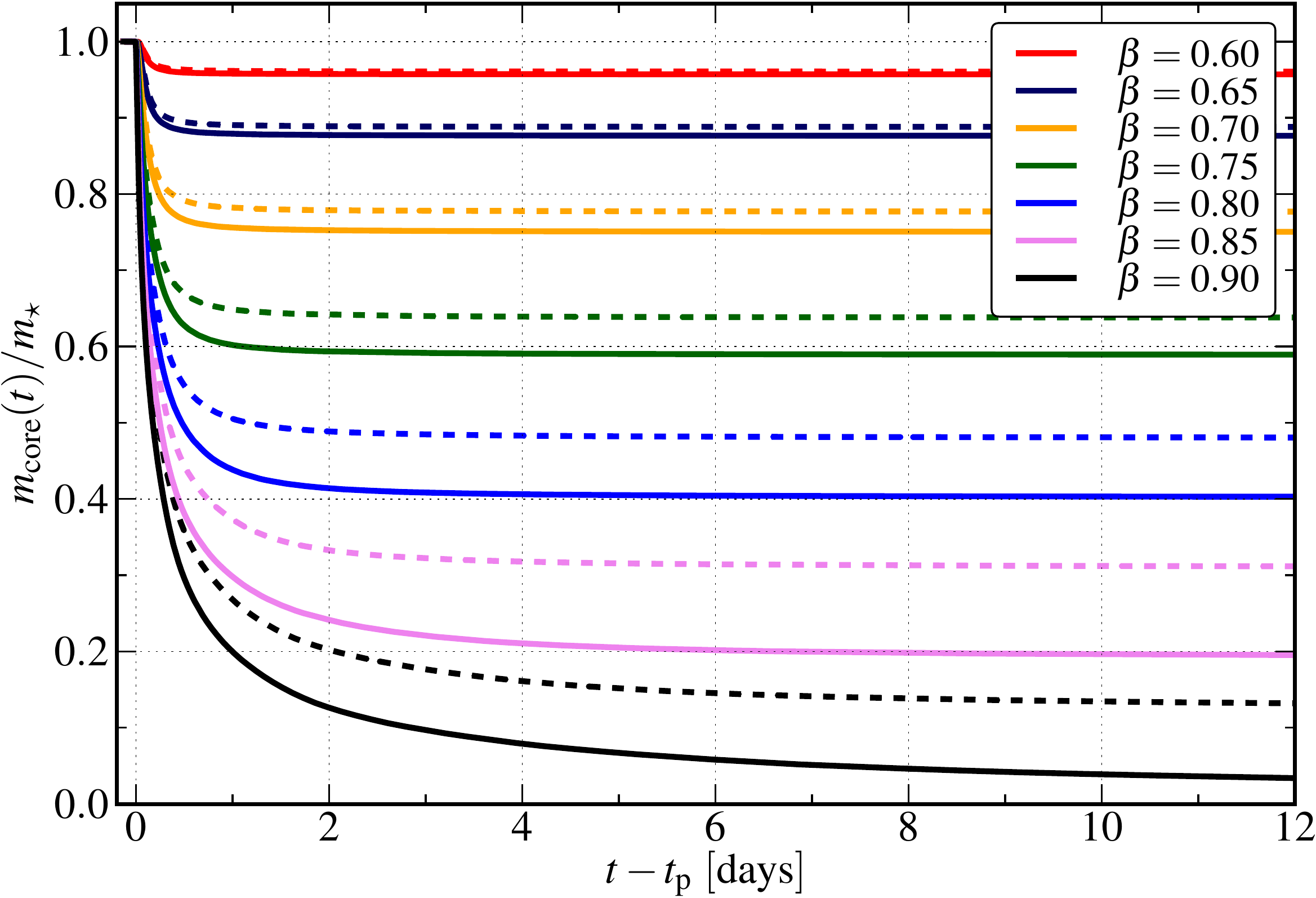}
\hspace*{5mm}
\includegraphics[width=.48\linewidth]{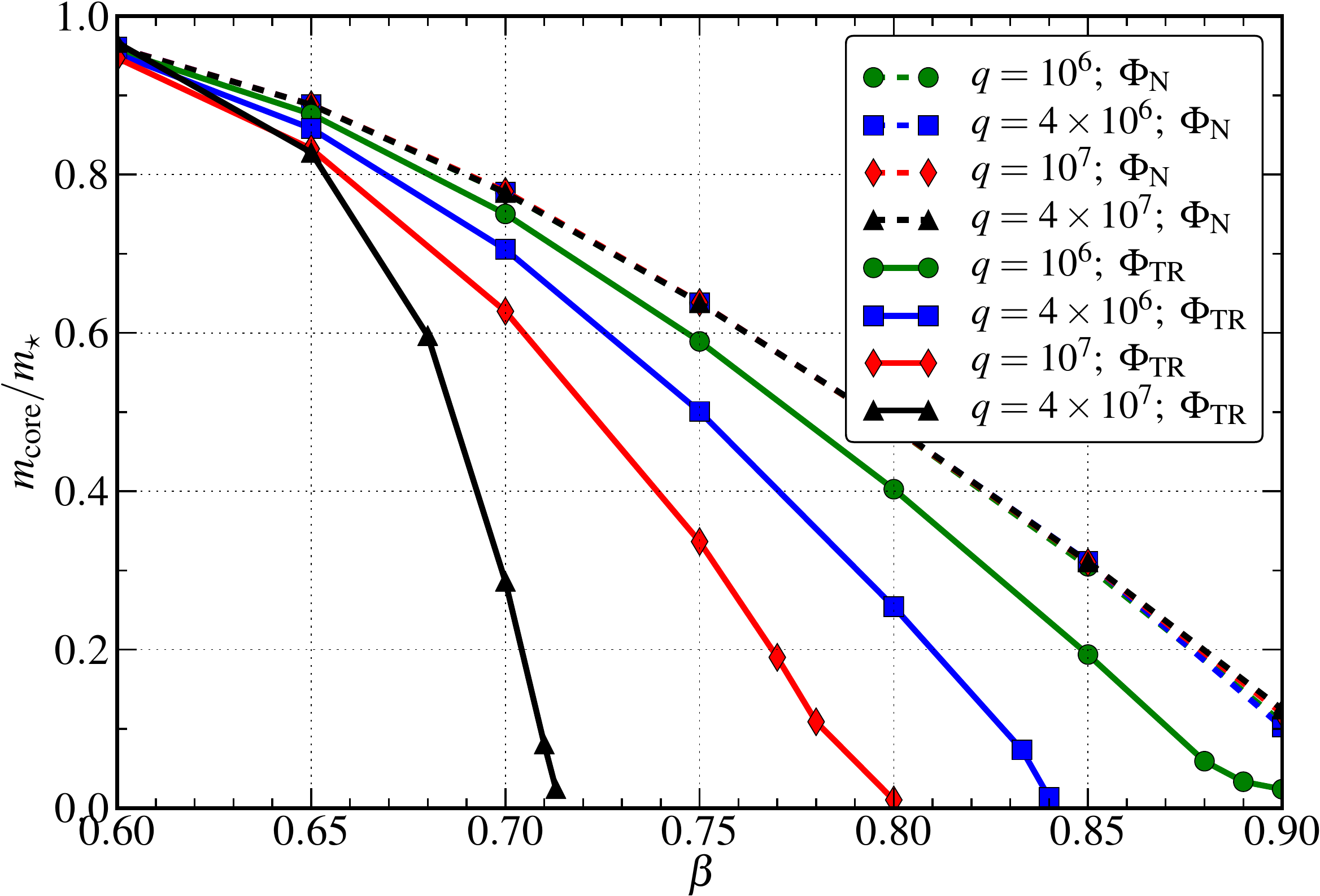}
\caption{\textit{Left panel}. Evolution of the self-bound mass fraction in simulations using 
Newtonian (dashed lines) and \citetalias{tejeda13} (solid lines) potentials.
\textit{Right panel}. Final self-bound mass fraction in Newtonian (dashed lines)
and \citetalias{tejeda13} (solid lines) simulations. The points corresponding
to Newtonian runs are esentially overlapping for all values of $\beta$, while
those from pseudo-relativistic simulations show a strong dependence on $q$:
the larger $q$ is, the smaller the critical $\beta$ necessary for complete
disruption is.}
\label{fig:mbound}
\end{figure*}

The self-bound mass fraction evolves during the disruption process, and for
a partial disruption it will decrease from $1$ (before disruption, when the entire star
is self-bound) to the final value $m_{\rm core}/m_\star$. In Fig.~\ref{fig:mbound} (left panel) we
present the time-evolution of $m_{\rm core}$ for the $q=10^6$ simulations. 
The disruption is stronger in relativistic encounters, with the
deviation of the self-bound mass fraction from the Newtonian case increasing with $\beta$, 
from $\sim$~few percent ($\beta=0.6$) to $\sim 100$~percent ($\beta=0.9$).

Fig.~\ref{fig:mbound} (right panel) shows the final self-bound mass fraction
as a function of $\beta$, for various ratios $q$ and for both potentials.
We observe that with the Newtonian potential
stars are partially disrupted in the range $0.6 \lesssim \beta \lesssim 0.9$ regardless of $q$. 
This result is general for all Newtonian disruptions of $\gamma=5/3$ polytropes, and agrees with the
numerical findings of e.g. \citet{guillochon13} ($\beta_{\rm d}$ necessary for
complete disruption equal to 0.9).
On the other hand, for any given $\beta$ 
the discrepancy between Newtonian and \citetalias{tejeda13} simulations increases 
drastically for larger black hole masses, as $\rs$ becomes comparable to $\rp$
and relativistic terms in the tidal tensor close to periapsis can no longer be ignored  
(see Appendix~\ref{sec:app}).

\subsection{Kick velocity}
\begin{figure*}
\centering
\includegraphics[width=.48\linewidth]{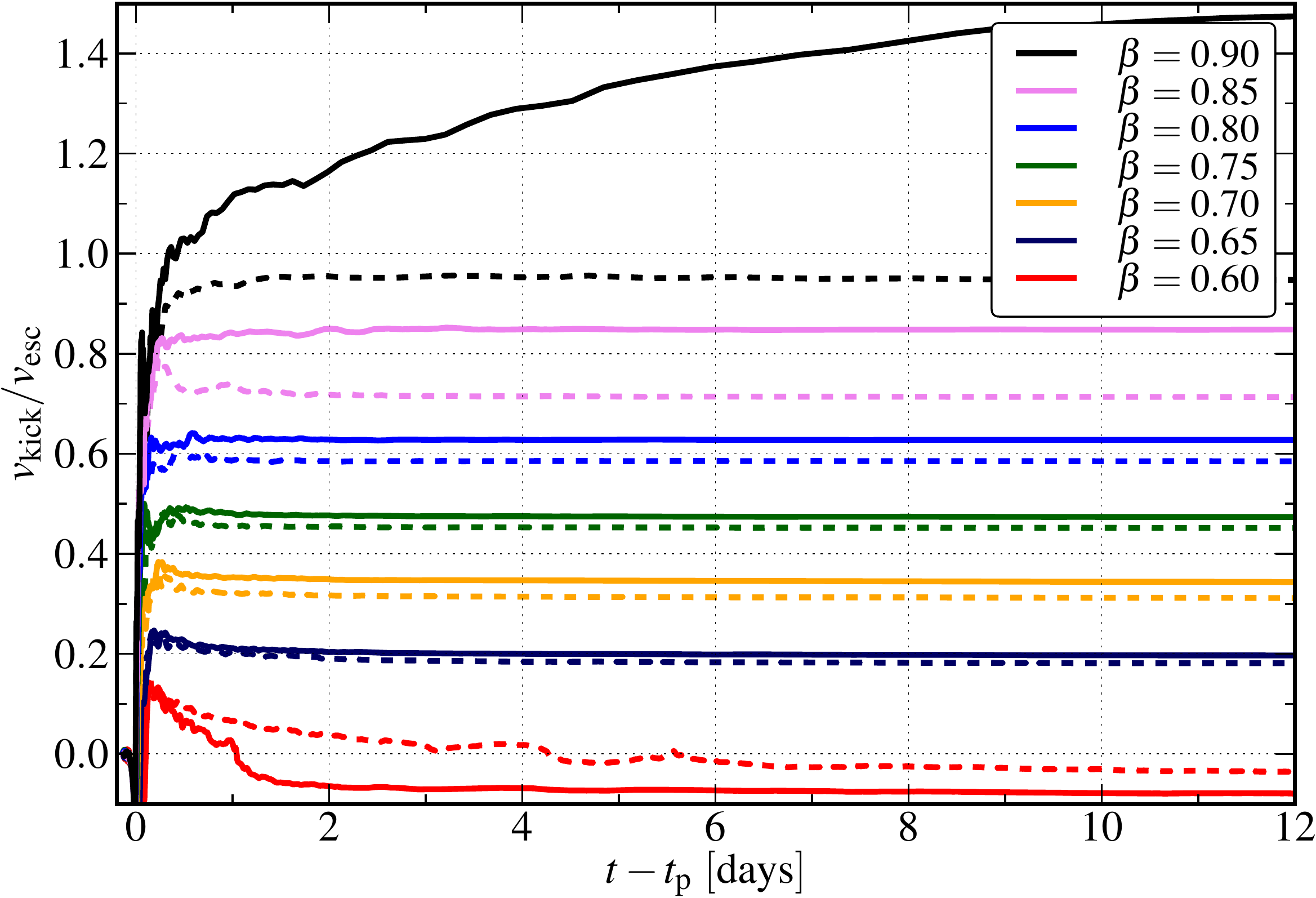}
\hspace*{5mm}
\includegraphics[width=.48\linewidth]{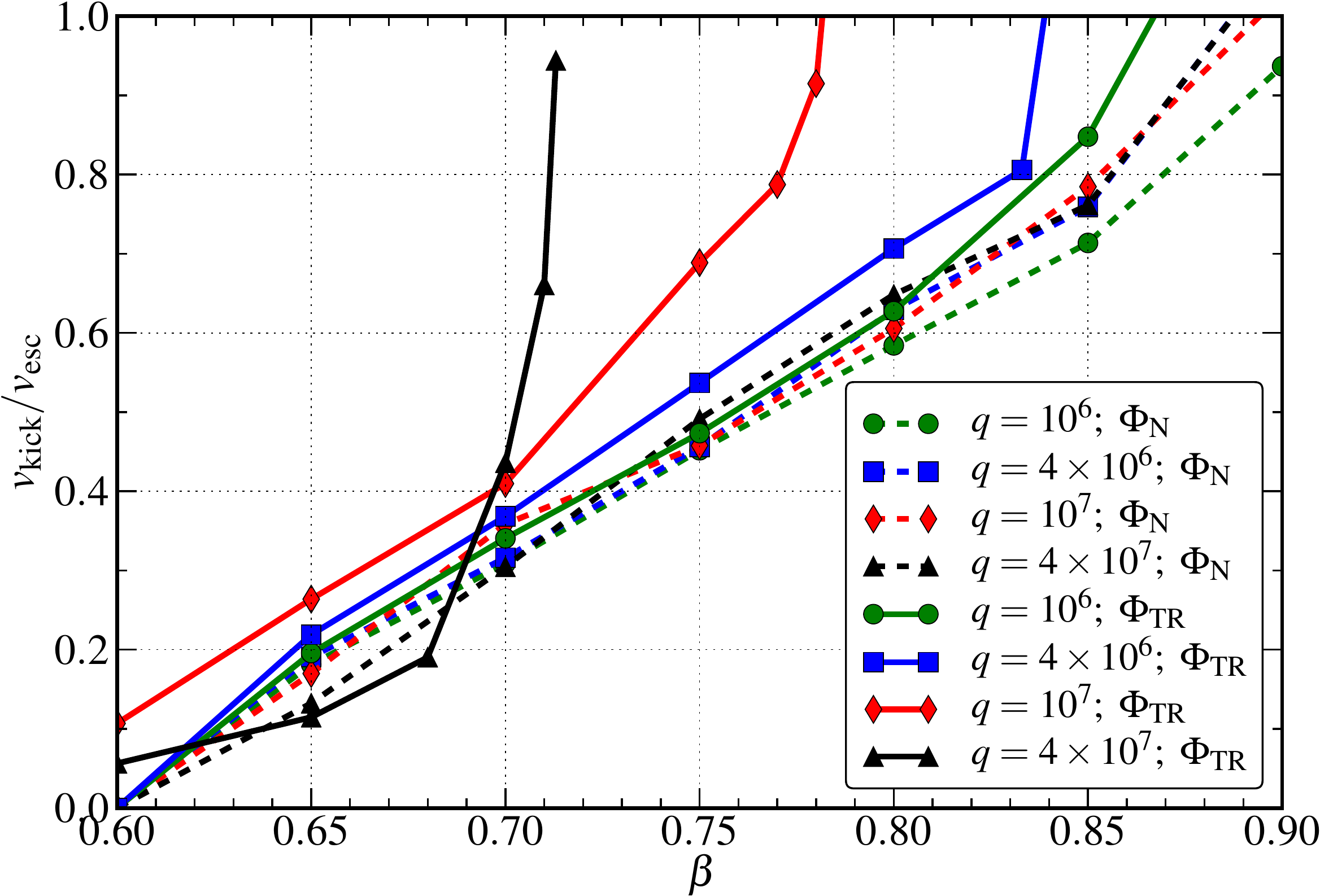}
\caption{\textit{Left panel}. Evolution of the kick velocity of the self-bound remnant in Newtonian 
(dashed lines) and \citetalias{tejeda13} (solid lines) simulations. The kick velocity is 
normalized by the surface escape velocity of a solar-type star, $\approx 617$ km/s. The lines show a 
moving average of the data points in order to smooth out fluctuations
during the disruption process (before the surviving remnant becomes relaxed, i.e. 
approximately during $t=0$ and $t=1$).
\textit{Right panel}. Kick velocity of the self-bound remnant at infinity, in Newtonian (dashed lines)
and \citetalias{tejeda13} (solid lines) simulations. The solid
black line ($q=4\times 10^7$, $\Phi=\Phi_{\rm TR}$) stands out, as for
low values of $\beta$ it suprisingly falls both below the runs
with smaller $q$, and below the Newtonian run. We have repeated this
set of simulations with higher resolution ($10^6$ SPH particles), but
the results were very similar. Since these are highly relativistic
encounters, further studies with even higher resolution and an exact
relativistic treatment of the black hole gravity are probably required
in order to arrive at a definitive answer..}
\label{fig:vbound}
\end{figure*}

The kick velocity is computed from the increase in specific orbital energy 
($\epsilon-\epsilon_0$) of the self-bound core using \eq{eq:vkick}.
In Fig.~\ref{fig:vbound} we present $v_{\rm kick}$ both as a function of time for the $q=10^6$ simulations
(left panel), and its final value at the end of the simulation (right panel).
The kick velocities are always scaled to
the surface escape velocity of the original star, $\sqrt{2GM_\odot/R_\odot}\approx 617$~km/s,
as they are expected to be comparable on theoretical grounds \citep[see][]{manukian13}.

We initially place the star at 5~$\rt$, and observe that numerical deviations from $\epsilon_0=0$ 
(i.e., parabolic trajectory) 
of the initial specific energy increases with $q$, and is larger for the \citetalias{tejeda13} potential.
For the most extreme case ($q=4\times 10^7$, $\beta=0.9$, $\Phi_{\rm TR}$), 
the initial energy was of the order of $\sim 10^{-8}$~c$^2$
(which is to be compared with the typical ``kick'' specific energy, $\sim 10^{-6}$~c$^2$), 
while for the other simulations it was between one and three orders of magnitude smaller 
(for the \citetalias{tejeda13} and Newtonian potentials, respectively).
Nevertheless, there is a clear distinction between the Newtonian runs, where
the kick is fairly independent of $q$, and the \citetalias{tejeda13} runs, where the curves exhibit
an asymptotic behaviour limited by progressively smaller $\beta$'s with larger $q$'s.

The time evolution of $\epsilon$ and -- consequently -- of $v_{\rm kick}$
exhibits significant oscillations during the actual disruption ($\sim$ hours from 
periapsis passage) that appear as wiggles in Fig.~\ref{fig:vbound}.
The explanation is likely related to the complicated hydrodynamic effects that take place 
in that short period of time.
Due to the extreme compression of the star, shock waves travel throughout the star and
transfer significant energy and angular momentum between the particles, and until the energies
become frozen-in, the average specific energy of the bound core keeps 
oscillating.

\begin{figure}
\centering
\includegraphics[width=\linewidth]{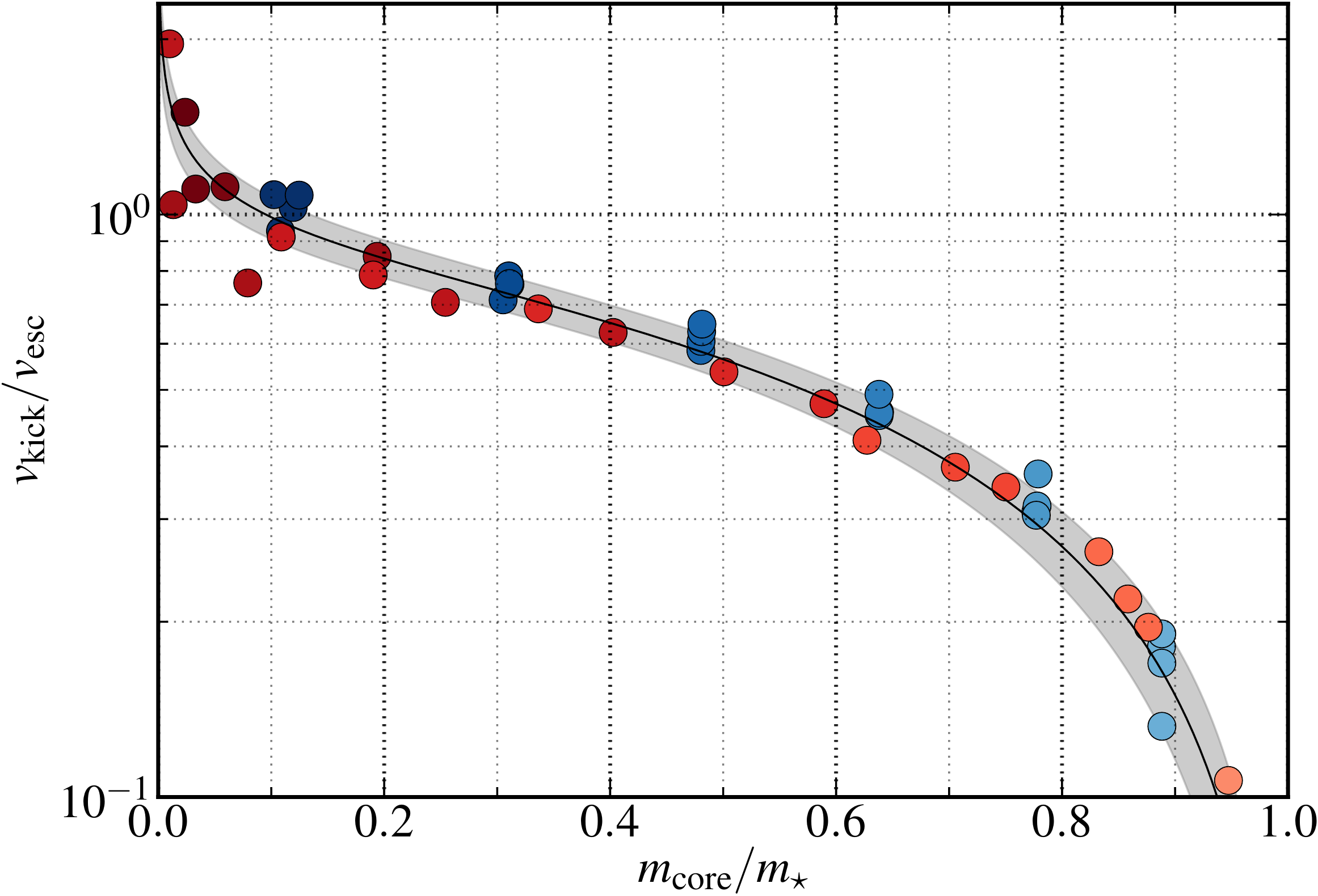}
\caption{The kick velocity $v_{\rm kick}$ of the self-bound remnant shows 
a remarkable dependency on the the self-bound mass fraction 
in both Newtonian (blue points) and \citetalias{tejeda13} (red points) 
simulations. We fitted a truncated power law of the form
$y=Ax^{-\alpha}(1-x^2)^\beta$, with the fit parameters
$A=0.634726$, $\alpha=0.196598$, $\beta=0.882387$. The shaded gray area
around the black fit line represents the $1\sigma$ deviation from the fit.
Darker points correspond to higher values of $\beta$.
}
\label{fig:ofbeta-mv}
\end{figure}

Fig.~\ref{fig:ofbeta-mv} shows the kick velocity as a function of the self-bound mass 
fraction $m_{\rm core}$. We notice that in contrast to the findings of \citetalias{manukian13}, 
in some simulations the kick velocity exceeds the escape velocity 
of the initial star, but only when the value of $m_{\rm core}$ is sufficiently small. 
Indeed, $v_{\rm kick}(m_{\rm core})$ seems to be a 
monotonic function that asymptotically approaches $0$ for $m_{\rm core}\rightarrow m_\star$
and $+\infty$ for $m_{\rm core}\rightarrow 0$.

\begin{figure*}
\centering
\includegraphics[width=.48\linewidth]{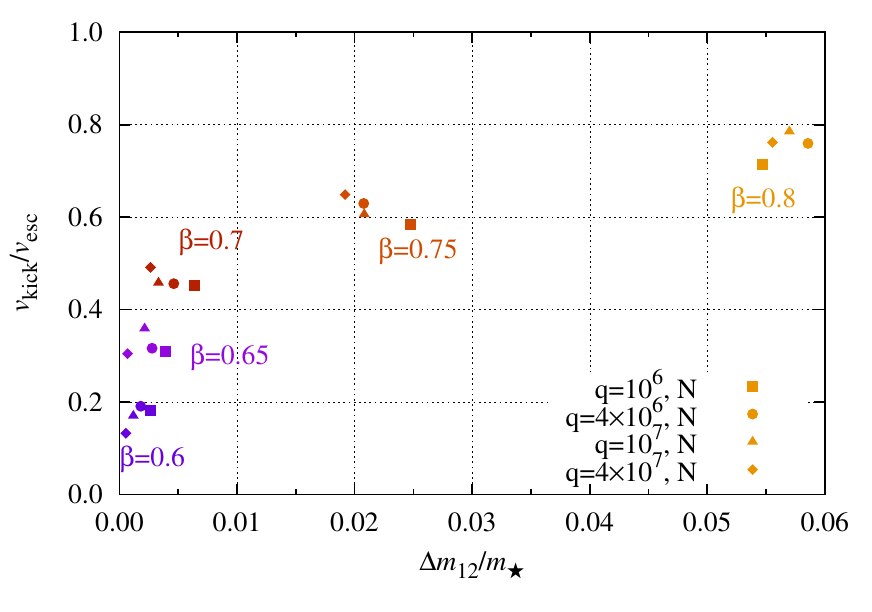}
\hspace*{5mm}
\includegraphics[width=.48\linewidth]{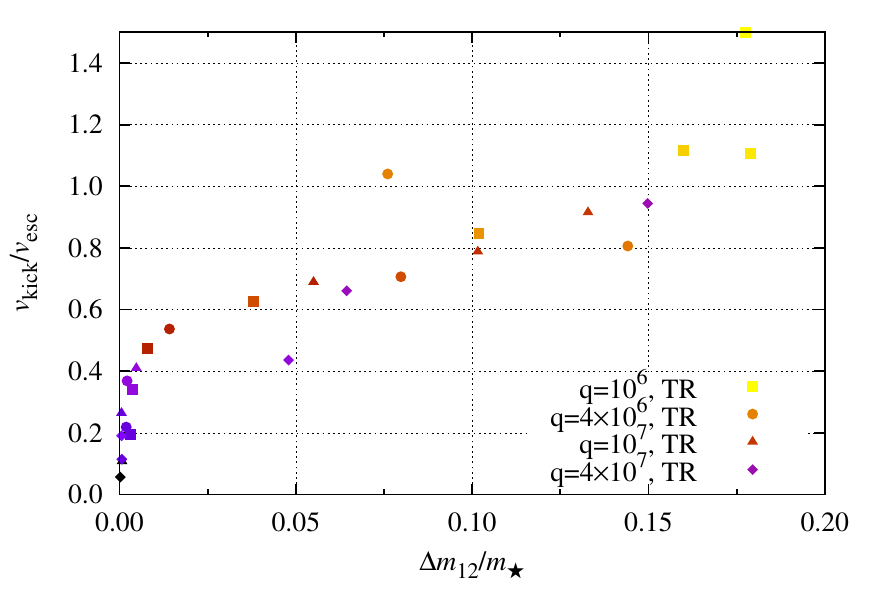}
\caption{Kick velocity of the self-bound remnant as a function of the mass difference $\Delta m_{12}$ between the tidal tails in Newtonian (\textit{left panel}) and \citetalias{tejeda13} (\textit{right panel}) simulations. The value of $\beta$ is colour-coded.
The relativistic results do not exhibit such a strong relation between $\beta$, $\Delta m_{12}$, and $v_{\rm kick}$ as in the case of the Newtonian potential, but they still obey the general trend 
($v_{\rm kick}$ increases with $\Delta m_{12}$) and even the shape of the function.}
\label{fig:ofbeta-mdif}
\end{figure*}

Fig.~\ref{fig:ofbeta-mdif} shows the kick velocity as a function of 
the mass difference between the two tidal tails,
$\Delta m_{12}$ (Fig.~\ref{fig:ofbeta-mdif}, left and right panels, for Newtonian
and \citetalias{tejeda13} potentials, respectively).
The two plots exhibit similar behaviours ($v_{\rm kick}$ increases with $\Delta m_{12}$),
but in the relativistic simulations the degeneracy in $q$ is broken (i.e., the data points
for a given $\beta$ are not clustered together irrespective of $q$),
since the relativistic kicks are sensitive to both $\beta$ and $q$.

\begin{figure}
\centering
\includegraphics[width=\linewidth]{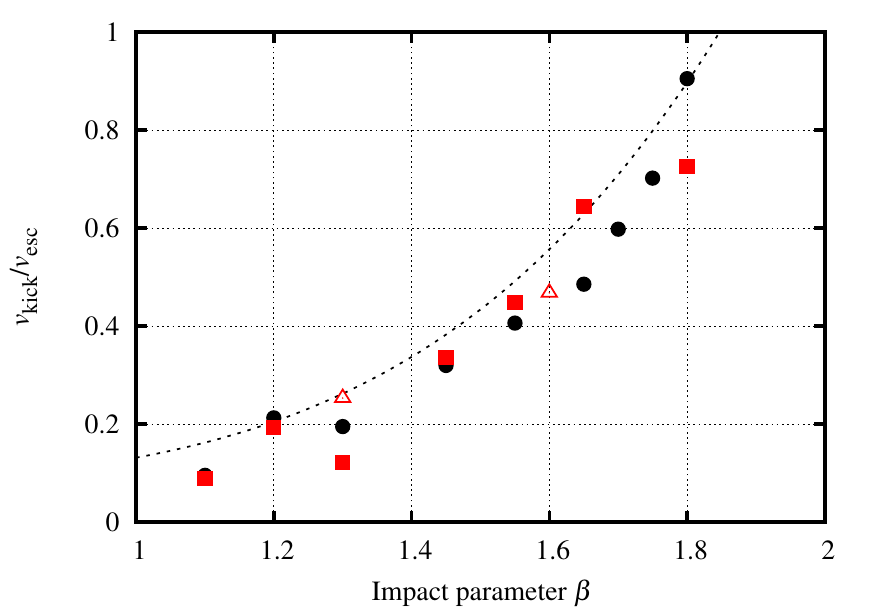}
\caption{The results of our SPH simulations of $\gamma=4/3$
  polytropes, using $N_{\rm part}=10^5$ in Newtonian (black
circles) and \citetalias{tejeda13} (red squares) potentials, 
and using $N_{\rm part}=10^6$ with the TR potential (red triangles), 
as compared to the fit line given by {\MGRO} 
for their $\gamma=4/3$ simulations. Our runs use the same
values for $q$ as in their paper ($10^3$ to $10^6$), 
and for this reason relativistic effects are small.}
\label{fig:fit43}
\end{figure}

We also show $v_{\rm kick}$ as a function of $\beta$
for the $\gamma=4/3$ simulations (Fig.~\ref{fig:fit43}), together with the fit line
given by \citetalias{manukian13}.
These simulations show reasonable
similarity to the results of \citetalias{manukian13}, and -- as expected -- for such small values of $q$
($10^3$ to $10^6$) there is little difference between Newtonian and \citetalias{tejeda13} simulations.
The data points from our simulations give slightly smaller kicks than the fit,
but the general trend (and critical $\beta$) are nevertheless recovered.

\subsection{Error estimation and resolution dependence}
\begin{figure*}
\centering
\includegraphics[width=\linewidth]{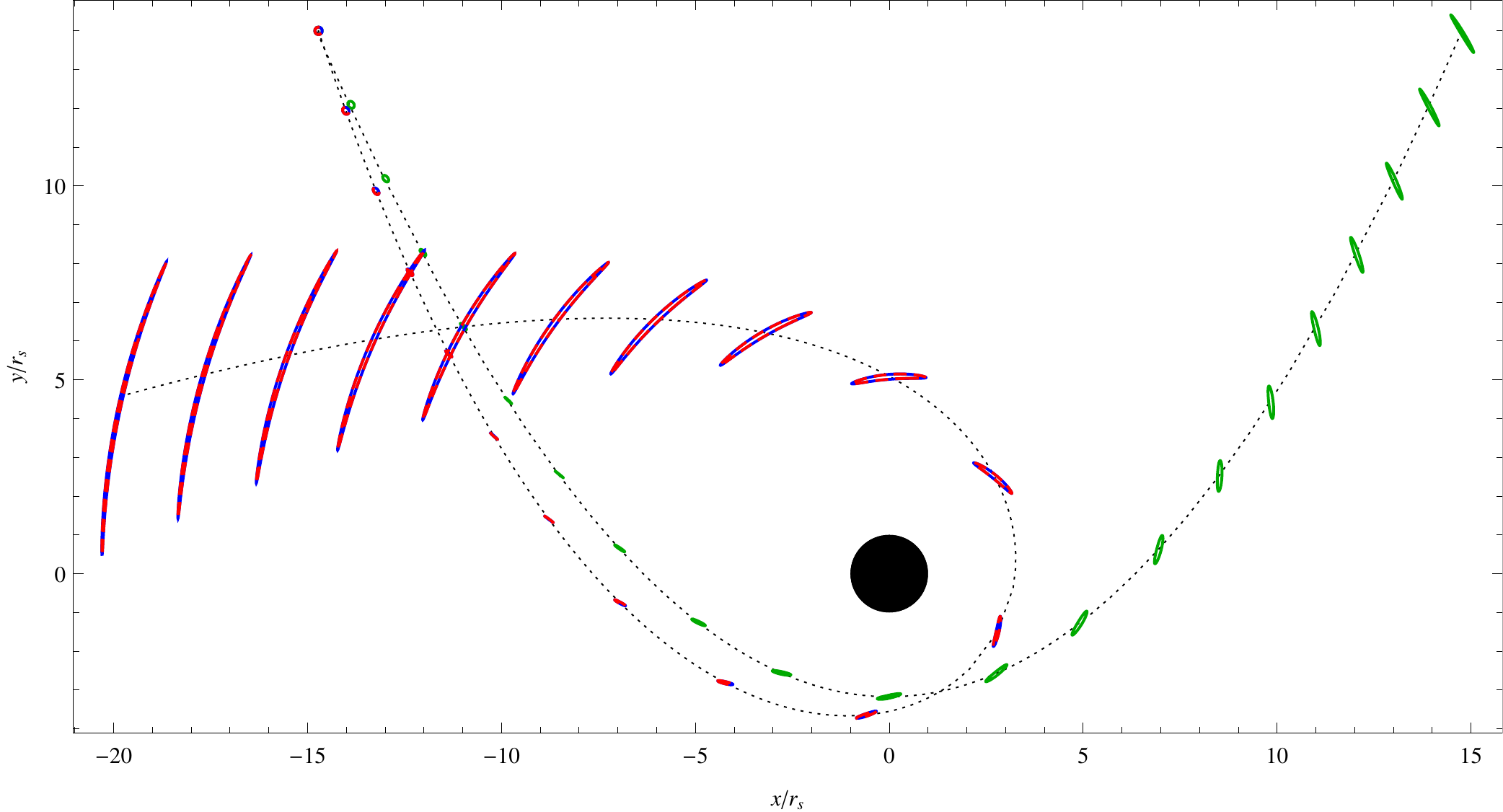}
\caption{Visual representation of the tidal deformation experienced by a ring
of particles of radius $4\,R_\odot$ (representing a star disrupted on a parabolic
orbit with $\beta=1.6$ by a SMBH with $q=10^7$) 
in Schwarzschild spacetime (red), the \citetalias{tejeda13} potential (blue), and
the Newtonian potential (green). The rings in Schwarzschild spacetime and the TR potential 
are indistinguishable from each other by eye, which is an indication that $\Phi_{\rm TR}$
reproduces the Schwarzschild tidal tensor to a very good accuracy (see Appendix~\ref{sec:app} for 
a further discussion). Unlike in the Newtonian case, where the shape of the rings would not change significantly after they exit the tidal radius, the other two rings become increasingly
deformed due to the different periapsis shifts of their component particles (the TR potential reproduces
the periapsis shift exactly).}
\label{fig:TidalDeformation}
\end{figure*}

\begin{figure*}
\includegraphics[width=.48\linewidth]{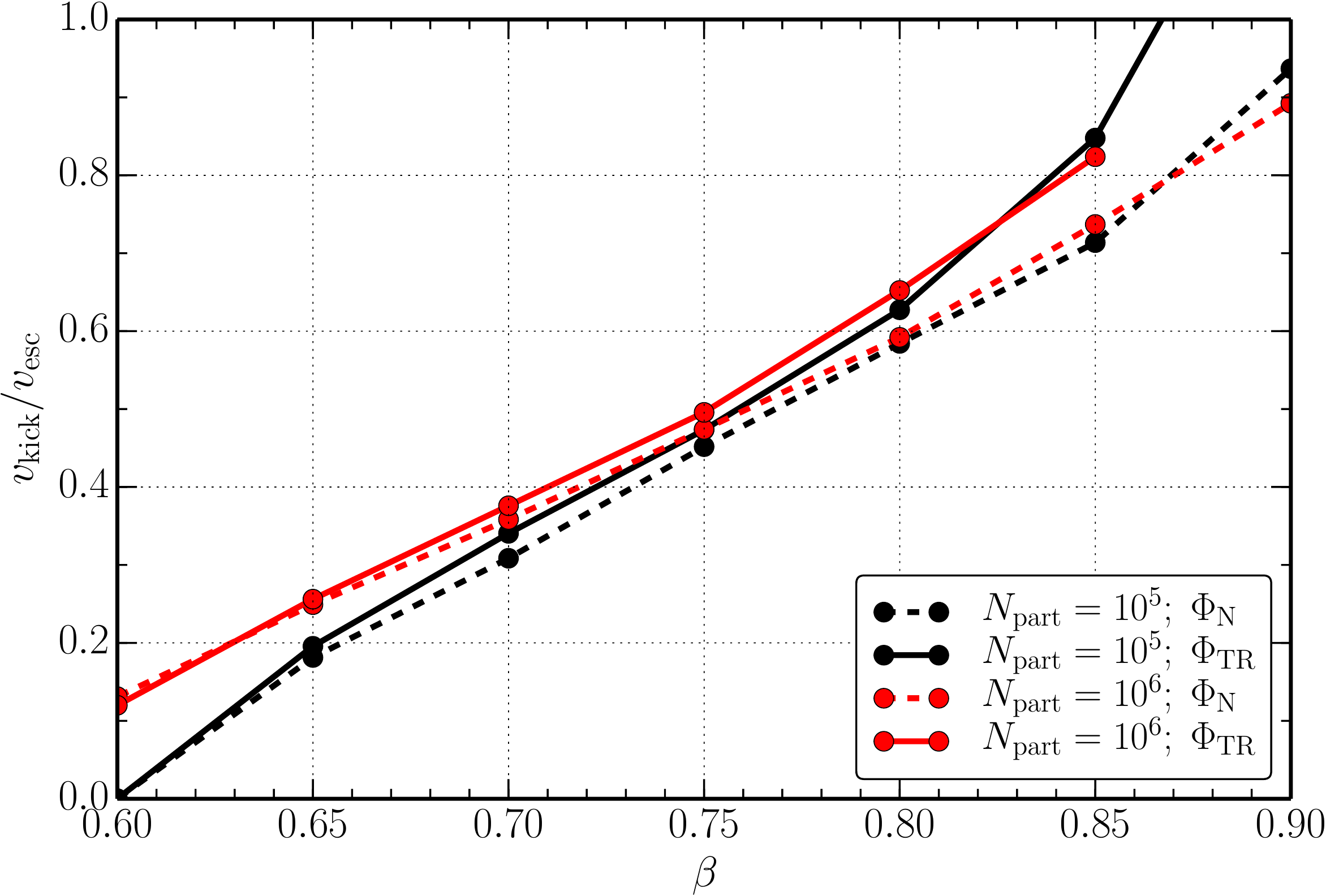}
\hspace*{5mm}
\includegraphics[width=.48\linewidth]{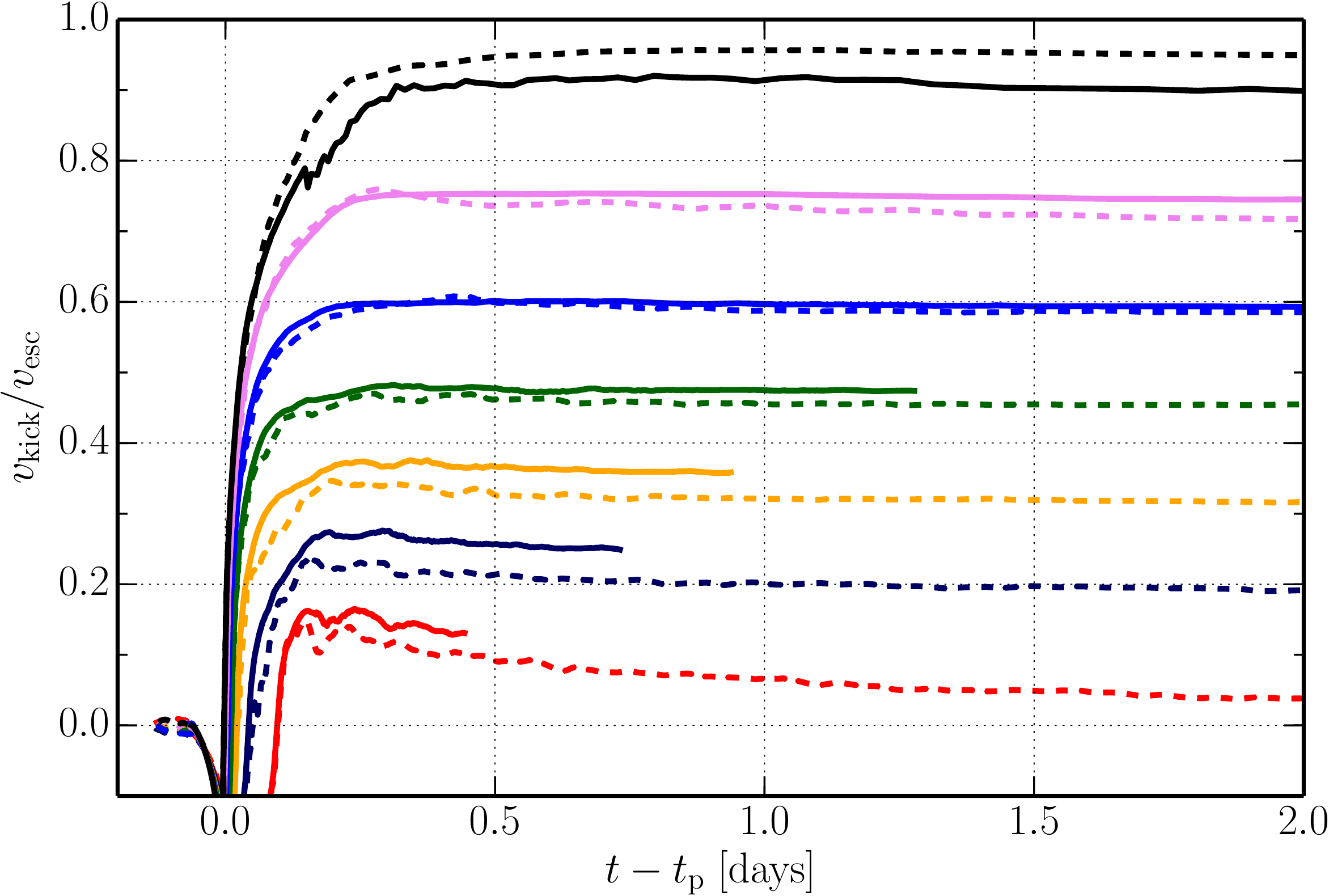}
\caption{\textit{Left panel}. Kick velocity as a function of impact
  parameter $\beta$, obtained with lower ($N_{\rm
    part}=10^5$; black lines) and higher ($N_{\rm part}=10^6$; red lines) resolution
  simulations, representing both $\Phi_{\rm N}$ (dashed lines) and $\Phi_{\rm
    TR}$ (solid lines) encounters.
  Resolution appears to be critical at low $\beta$ ($\sim 0.6$), where the mass asymetry of
  the two tails must be properly resolved in order to extract an
  accurate kick velocity (nevertheless, at both resolutions there is a
  clear trend of higher kick velocities in the relativistic case,
  compared to the Newtonian case). 
For the central part of the range,
  simulations with the two resolutions agree to within $\sim 1$
  percent. At large $\beta$, close to the critical impact parameter
  ($\sim 0.9$), the two results agree for $\Phi=\Phi_{\rm N}$, but for
  the TR encounter the $10^6$ particles simulation results in no
  surviving core, while the $10^5$ particles simulation still yields a small
  self-bound mass that acquires all the energy, resulting in a very
  large kick. \textit{Right panel}. Time evolution of the kick velocity
  for $q=10^6$, $\Phi=\Phi_{\rm
    TR}$, and $N_{\rm part}=10^5$ (dashed lines) and $N_{\rm
    part}=10^6$ (solid lines). The different colours represent
  different values of $\beta$. The best agreement between the kick
  velocities is achieved in the region of intermediate $\beta$, 
  where both the mass asymmetry of the two tails and the
  surviving core are resolved in the lower-resolution runs. 
  Since a significant fraction of the star survives for low
  $\beta$'s, it is very expensive to run these simulations with high
  resolution, and therefore they were stopped at an earlier physical time
  than the high $\beta$ simulations.}
\label{fig:hires}
\end{figure*}

Since in this paper we are concerned with parabolic encounters, 
we have calculated the \citetalias{tejeda13} acceleration for particles on parabolic orbits.
It turns out that the TR potential reproduces parabolic orbits exactly 
(see \citealp{tejeda13}; in our simulations, these would be the orbits of the centre of mass).
In Fig.~\ref{fig:TidalDeformation} we show the tidal deformation experienced
by test particles placed on a ring of radius $4\,R_\odot$ with the centre following
a parabolic orbit with $\beta=1.6$ around a $10^7~M_\odot$ black hole. The 
deformation of the ring provides a visual representation of the
effects of the tidal tensor, and by eye it appears to be identical 
in Schwarzschild and $\Phi_{\rm TR}$. 
In Appendix~\ref{sec:app} we calculate in detail the 
components and the eigenvalues of the tidal tensor in Schwarzschild and
both Newtonian and \citetalias{tejeda13} potentials, and show that the eigenvalues $\lambda_2$ and
$\lambda_3$ (corresponding to compression in the orbital plane and perpendicular
to it, respectively) provided by the \citetalias{tejeda13} potential for parabolic orbits are 
the same as in Schwarzschild (with $\lambda_3$ having an identical expression,
and $\lambda_2$ having a slightly different expression that still gives the
same result within machine precision for our range of parameters), 
while the maximum relative error in 
$\lambda_1$ (corresponding to expansion in the radial direction) 
is 6.4\%, occurring at periapsis for a very narrow range of $\beta$'s.
In general, the average error in $\lambda_1$ at periapsis is of order $\sim$ a
percent, and quickly drops farther away from the black hole. 
We also show that both the 
Newtonian and the \citetalias{tejeda13} potentials \textit{always underestimate} the values of 
$\lambda_1$ and $\lambda_2$, which means that all disruptions will be slightly stronger
in the Schwarzschild spacetime, leaving a smaller self-bound core and resulting
in even larger velocity kicks. Therefore, our results are robust lower limits on the true relativistic effects.\\

\noindent We have also performed a number of test simulations with higher
resolution. For $q=10^6$, we ran all simulations with both $10^5$ and
  $10^6$ SPH particles; a comparison of the resulting kick velocities is presented in Fig.~\ref{fig:hires}.
The most striking observation is that resolution is crucial at low
$\beta$ ($\sim 0.6$), where the mass asymmetry between the two tidal tails,
which drives the kick, is extremely small, $\lesssim
10^{-2}~m_\star$. This mass loss must be properly resolved in order
for the kick effect to be accurately captured; we
observe the low-resolution simulations greatly
underestimate the kick for $\beta=0.6$, due to more energy being
deposited into the oscillation modes of the star. Resolution is
also important, though not as crucial, at high $\beta$ ($\sim 0.9$),
where the surviving core itself is of the order of $\sim 10^{-2}~m_\star$, and
must also be properly resolved; we observe that here the discrepancy
between $\Phi_{\rm N}$ and $\Phi_{\rm TR}$ is less extreme for the higher
resolution simulations, but this is simply because $\Phi_{\rm TR}$ leaves a
small surviving core, which is not resolved by the low-resolution 
simulations. 
For intermediate values of $\beta$, where both
the core and the mass difference of the tails are a generous
fraction of the initial stellar mass, lower-resolution simulations 
agree to a reasonable accuracy ($\sim$ few percent) with the results 
obtained with the $10^6$ particles simulations.

For $q=10^7$, we ran the simulation with $\beta=0.7$ with various 
numbers of SPH particles $N_{\rm part}$ and tree opening angles $\theta$: 
for $\Phi_{\rm N}$, $N_{\rm part}=10^5$ and 
$N_{\rm part}=10^6$ (both with $\theta=0.5$),
and for $\Phi_{\rm TR}$, $N_{\rm part}=10^5$ ($\theta=0.2$, $0.5$),
and $N_{\rm part}=10^6$ ($\theta=0.5$).
We observe that improving the force accuracy above $\theta=0.5$ does
not have a significant impact on the final kick velocity 
(at most $\sim$ one percent), 
while increasing the number of particles ten times does change it by up to 
$\sim$ 10 percent,
for both potentials. In addition, the evolution of $\epsilon_{\rm kick}$ during the 
disruption process is highly dependent on the resolution, 
with larger wiggles in the high-resolution simulations, 
presumably due to the fact that shocks are better resolved and
therefore less dissipative, and there is significantly more energy transfer between the particles.
Still, the difference between the kick velocities of the Newtonian
and \citetalias{tejeda13} potentials is consistently and significantly
larger than the variations that appear when changing the accuracy
parameters for the same potential.

For $\gamma=4/3$, we ran two additional simulations with $N_{\rm
    part}=10^6$ and $\Phi=\Phi_{\rm TR}$, using $q=10^4$, $\beta=1.6$ and $q=10^6$,
  $\beta=1.3$. The kicks obtained in these higher-resolution
  simulations are shown in Fig.~\ref{fig:fit43} with red squares, and are in
  good agreement with both the fit given by MGRO and the rest of the
  points from our simulations.

\section{Discussion}\label{sec:discussion}
We have confirmed by a different numerical method (SPH vs. the
adaptive mesh refinement code FLASH that is primarily used by \citetalias{manukian13}) 
that all Newtonian simulations with the same impact parameter $\beta$ 
produce similar self-bound remnants (in mass, specific energy, 
and kick velocity), confirming earlier conclusions by \citet{manukian13} that $v_{\rm kick}$ is virtually independent of the mass ratio $q$.

We compared these calculations with simulations that use a generalized potential that
accurately captures the dynamics around a Schwarzschild black hole.
Unlike Newtonian tidal disruptions, we observe that relativistic disruptions
are no longer fully described by the parameter $\beta$.
Instead, relativistic effects related to the tidal tensor become important 
when the periapsis distance
is comparable to the Schwarzschild radius, i.e. they depend on the ratio
\begin{equation}\label{eq:def tildebeta ext}
\Theta \equiv \frac{\rs}{\rp}
       = \beta\frac{m_\star}{r_\star} q^{2/3} \frac{2G}{c^2} 
       \approx 0.2 \times \frac{\beta}{5}
                  \left(\frac{r_\star}{R_\odot}\right)^{-1}
                  \frac{m_\star}{M_\odot}
                  \left(\frac{q}{10^6}\right)^{2/3}.
\end{equation}

For a given mass ratio $q$, this can be interpreted as a dependence
on $\beta$ (e.g., for $q=10^6$, relativistic effects become important when
$\beta \gtrsim 5$, as noticed by e.g. \citealt{laguna93}), 
which will however change with $q$ (for $q=10^7$ relativistic
effects are extremely important even for $\beta\lesssim 1$), since the gravitational
radius and the tidal radius have different dependencies on $M_{\rm bh}$.
In general, we observe that relativistic effects can be ignored for
$\Theta \lesssim 10^{-2}$ (where the relativistic and the Newtonian tidal tensor are essentially the same,
and relativistic effects such as periapsis shift are negligible on the time scale of the disruption), 
but tend to dominate the outcome of the encounter 
for $\Theta \gtrsim 0.1$. 
These thresholds can also be observed in Fig.~\ref{f1} in the Appendix, 
where we see that in a typical 
tidal disruption the eigenvalues of the tidal tensor are virtually identical
in Newton and Schwarzschild for $r\gtrsim 30\,\rs$, and start to differ 
significantly from each other for $r\lesssim 10\,\rs$ (the exact values will of course
depend on the parameters of the encounter).

We therefore expect relativistic effects on the kick velocity
of turbo-velocity stars to be significant for larger black hole masses.
In particular, since these objects will only result from partial disruptions ($0.6 \lesssim
\beta \lesssim 0.9$, with the more pronounced velocity kicks at the higher end of
this range), relativistic effects should dominate for $\Theta \gtrsim 0.1$,
or, from \eq{eq:def tildebeta ext}, for
\begin{equation}
q \gtrsim 4.2\times 10^6 \left(\frac{m_\star}{M_\odot}\right)^{-3/2}\left(\frac{r_\star}{R_\odot}\right)^{3/2}.
\end{equation}
This means that for a solar-type star
disrupted in our Galactic centre relativistic effects may be important. In Fig.~\ref{fig:vbound} we
observe that for $q=4\times 10^6$ the \citetalias{tejeda13} potential will yield the same kick velocity 
as the Newtonian potential at a $\sim 5\%$ smaller $\beta$.
Since tidal disruption rates scale with $\sim\beta^{-2}$
we estimate that approximately $\sim 10\%$ more stars will have a given kick velocity compared 
to a Newtonian estimation.

Previous estimations by \citet{kesden12a,kesden12b} predict 
that the spin of the black hole may alter the spread in energy by
up to a factor of $\sim$ 2. If one were to anticipate the relativistic effects due
to the black hole spin, one would therefore expect a maximum of
$\sim$ 40\% increase in the kick velocity, depending on the spin and
orbit orientation, though a methodical study of such effects is left for subsequent
investigations.

To conclude, we have found that the critical $\beta$ necessary for the disruption of the star is highly
dependent on $q$ as long as $\Theta \gtrsim 10^{-2}$, with 
the star being completely disrupted at a $\beta$ of around 0.9 ($q=10^6$),
0.85 ($q=4\times 10^6$), 0.8 ($q=10^7$), 0.72 ($q=4\times 10^7$),
due to the proximity to the event horizon ($\Theta\approx 0.04$, 0.09,
0.16, and 0.36, respectively).
This implies that the higher $q$ is, the smaller
the span of $\beta$'s in which partial disruption will occur, 
and the steeper the dependency of the surviving core's mass, 
$m_{\rm core}$ on $\beta$ is (Fig.~\ref{fig:mbound}).
Since we have also shown that there is a very clear, monotonic relation between $m_{\rm core}$ 
and the kick velocity $v_{\rm kick}$ imparted to the core (Fig.~\ref{fig:ofbeta-mv}), 
we conclude that -- in Schwarzschild spacetime -- heavier
black holes are able to impart larger kick velocities without requiring very deep encounters.

%
%
\section*{Acknowledgements}
The simulations of this paper were in part performed on the facilities of the 
H\"{o}chstleistungsrechenzentrum Nord (HLRN) in Hannover, and at the 
PDC Centre for High Performance Computing (PDC-HPC) in Stockholm.
We thank John Miller for the careful reading of the manuscript, and
we acknowledge useful discussions on the topic of the paper with Enrico Ramirez-Ruiz.
In addition, we thank the referee, Tamara Bogdanovi\'c, for insightful
comments and very helpful suggestions.
The work of SR has been supported by the Swedish Research Council (VR) 
under grant 621-2012-4870.

%
%
\bibliographystyle{mn2e}
\bibliography{paper}

\appendix

%
%
\section{Derivation of the tidal gravitational field for the TR potential}
\label{sec:app}

In this appendix we provide explicit expressions for the tidal tensor
corresponding to the \citetalias{tejeda13} potential and quantify its
departure from the exact relativistic result in Schwarzschild
spacetime. The acceleration exerted on a given test particle under the
TR potential is given by \citep{tejeda13}
\begin{equation}
\ddot{x}_{i} =  -\frac{\G M_{\rm bh}x_{i}}{r^3}\left(1-\frac{\rs}{r}\right)^2 +
 \frac{\rs\,\dot{x}_{i}\,\dot{r}}{r(r-\rs)}
 - \frac{3}{2}\frac{\rs\,x_{i}\,\dot{\varphi}^2}{r},
 \label{e1} 
\end{equation}
where $x_i=\{x,\,y,\,z\}$ and $r$, $\dot{r}$ and $\dot{\varphi}$ should be taken as implicit functions of the Cartesian coordinates satisfying
\begin{gather}
r^2 = x^2+y^2+z^2,\\
r\,\dot{r} = x\,\dot{x}+y\,\dot{y}+z\,\dot{z},\\
r^4 \dot{\varphi}^2 = (x\,\dot{y}-y\,\dot{x})^2+(x\,\dot{z}-z\,\dot{x})^2+(z\,\dot{y}-y\,\dot{z})^2.
\end{gather}

\noindent We now consider a star approaching the central black hole. The centre of mass of the star is located at $\mathbfit{x}_0$ and follows, to a very good approximation, the trajectory of a free-falling test particle. We can compute the tidal forces acting on a fluid element located at a generic position $\mathbfit{x}$ within the star by taking the Taylor series expansion of \eq{e1} around $\mathbfit{x}_0$, i.e.
\begin{equation}
\ddot{x}_{i} =f_i(\mathbfit{x},\dot{\mathbfit{x}}) \simeq \left.f_i\right|_{(\mathbfit{x}_0,\,\dot{\mathbfit{x}}_0)} + 
\left(\mathbfit{x}- \mathbfit{x}_0\right)_j\left.\frac{\partial f_i}{\partial x_j}\right|_{(\mathbfit{x}_0,\,\dot{\mathbfit{x}}_0)} + 
\left(\dot{\mathbfit{x}}-\dot{\mathbfit{x}}_0\right)_j\left.\frac{\partial f_i}{\partial \dot{x}_j}\right|_{(\mathbfit{x}_0,\,\dot{\mathbfit{x}}_0)},
\label{e5}
\end{equation} where the Einstein summation convention is used.
Next we substitute $\mathbf{\xi}=\mathbfit{x}-\mathbfit{x}_0$ into \eq{e5} and find that, to first order in $\mathbf{\xi}$ and $\dot{\mathbf{\xi}}$, the acceleration acting on a fluid element due to the central black hole as seen from the centre of mass of the star is given by
\begin{equation}
\ddot{\xi}_{i} = \xi_{j} \left.C_{ij}\right|_{(\mathbfit{x}_0,\,\dot{\mathbfit{x}}_0)} 
+ \dot{\xi}_{j} \left.\widetilde{C}_{ij}\right|_{(\mathbfit{x}_0,\,\dot{\mathbfit{x}}_0)},
\label{e6}
\end{equation}
with the tidal tensors $C_{ij}$ and $\widetilde{C}_{ij}$ given by
\begin{align}
C_{ij} \equiv
\frac{\partial f_i}{\partial x_j}  = &
 -\frac{\G M_{\rm bh}}{r^3}\left(1-\frac{\rs}{r}\right)^2
\left[\delta_{ij} -\left(\frac{3\,r-5\,\rs}{r-\rs}\right)
\frac{x_i\,x_j}{r^2}\right]\nonumber\\
& +\frac{\rs\,\dot{x}_i\,\dot{x}_j}{r^2(r-\rs)}\left[
\delta_{ij} -\left(\frac{3\,r-2\,\rs}{r-\rs}\right)
\frac{\dot{r}\,x_j}{r\,\,\dot{x}_j}\right]\label{CTR}\\
& -\frac{3\,\rs}{2\,r}\dot{\varphi}^2
\left(\delta_{ij}-\frac{5\,x_i\,x_j}{r^2}\right)
 -\frac{3\,\rs\,x_i}{r^5}\left(x_j v^2-r\,\dot{r}\,\dot{x}_j\right),\nonumber\\
\widetilde{C}_{ij} \equiv 
\frac{\partial f_i}{\partial \dot{x}_j} = &\ 
\frac{\rs\left(r\,\dot{r}\,\delta_{ij}+x_j\,\dot{x}_i\right)}{r^2(r-\rs)}
 -\frac{3\,\rs\,x_i}{r^4}\left(r\,\dot{x}_j-\dot{r}\,x_j\right),
 \label{C2TR}
\end{align}
where $\delta_{ij}$ is the Kronecker delta symbol and $v^2 = \dot{x}^2+\dot{y}^2+\dot{z}^2$.

Now we would like to compare the results in \eqs{CTR} and \eqref{C2TR} with the corresponding exact relativistic expressions. It is customary to express the tidal tensor in Schwarzschild spacetime by adopting the so-called Fermi normal coordinates centred on the approaching star \citep[see e.g.][]{marck83,brassart10}. Nevertheless, for consistency with the approach that we have adopted here, we need to calculate the tidal field using Schwarzschild coordinates (i.e.~in the global reference frame centred on the black hole). The starting point is then the full-relativistic expression for the acceleration acting on a test particle in Schwarzschild spacetime
\begin{equation}
\ddot{x}_{i} =  -\frac{\G M_{\rm bh}x_{i}}{r^3}\left(1-\frac{\rs}{r}\right) + \frac{\rs\,\dot{x}_{i}\,\dot{r}}{r(r-\rs)}+
 \frac{\rs\,x_{i}\,\dot{r}^2}{2(r-\rs)r^2} - \frac{\rs\,x_{i}\,\dot{\varphi}^2}{r}.
 \label{eS1} 
\end{equation}
Following the same steps leading to \eq{e6}, it is found that the tidal tensors are now given by
\begin{align}
C^S_{ij} = &
-\frac{\G M_{\rm bh}}{r^3}\left(1-\frac{\rs}{r}\right)
\left[\delta_{ij} -\left(\frac{3\,r-4\,\rs}{r-\rs}\right)
\frac{x_i\,x_j}{r^2}\right]\nonumber\\
& +\frac{\rs\,\dot{x}_i\,\dot{x}_j}{r^2(r-\rs)}
\left[\delta_{ij}+
\frac{\dot{r}\,x_i}{r\,\dot{x}_i}-\left(\frac{3\,r-2\,\rs}{r-\rs}\right)
\frac{\dot{r}\,x_j}{r\,\dot{x}_j}\right] \\
& +\frac{\rs\,\dot{r}^2}{2\,r^2(r-\rs)}\left[
\delta_{ij} -\left(\frac{5\,r-4\,\rs}{r-\rs}\right)
\frac{x_i\,x_j}{r^2}\right]\nonumber\\
& -\frac{\rs}{r}\dot{\varphi}^2
\left(\delta_{ij}-\frac{5\,x_i\,x_j}{r^2}\right)
 -\frac{2\,\rs\,x_i}{r^5}\left(x_j v^2-r\,\dot{r}\,\dot{x}_j\right),\nonumber\\
 \widetilde{C}^S_{ij}  = &\ 
\frac{\rs\left(r^2\dot{r}\,\delta_{ij}+x_j\,x_i\,\dot{r} + r\,x_j\,\dot{x}_i\right) }{r^3(r-\rs)}
 -\frac{2\,\rs\,x_i}{r^4}\left(r\,\dot{x}_j-\dot{r}\,x_j\right),
\end{align}
where the superscript `S' is used to indicate that the quantity has been calculated in Schwarzschild spacetime.
\begin{figure}
\begin{center}
\includegraphics[width=\linewidth]{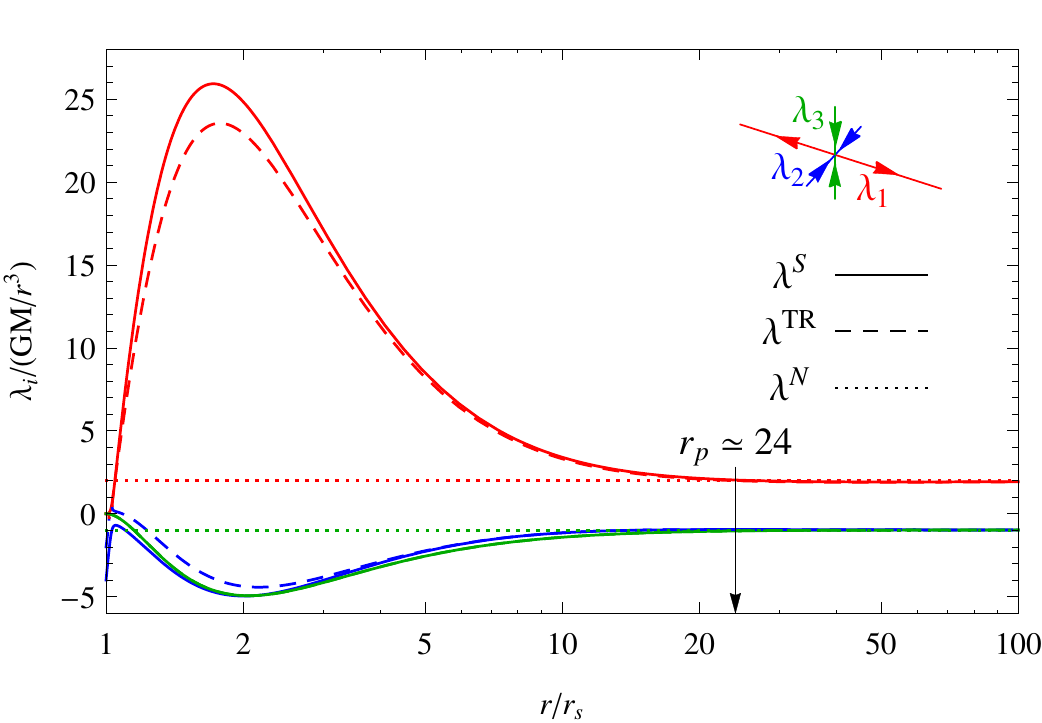}
\end{center}
\caption{Comparison of the eigenvalues of the tidal tensor as obtained in Schwarzschild spacetime (solid lines), for the \citetalias{tejeda13} potential (dashed lines), and for the Newtonian potential (dotted lines). In this case we have taken a parabolic encounter with $h=5\,\rs\,\cc$, corresponding to a $\beta \simeq 1$ encounter between a solar-type star with a $10^6\,M_{\odot}$ central black hole.}
\label{f1}
\end{figure}

The eigenvalues of the tidal tensor $C_{ij}$ give pertinent physical information about the amount of compression or expansion that the stellar matter experiences due to the black hole along the direction of the principal axes of the tidal tensor. These eigenvalues can be found by following the standard procedure of diagonalizing $C_{ij}$. In particular, if we choose a global reference frame XYZ such that the trajectory followed by the centre of mass of the star coincides with the XY plane (which is always possible due to the spherical symmetry of the present problem), the corresponding eigenvalues are given by
\begin{gather}
\lambda_{1,2} = \frac{1}{2}
\left[C_{11}+C_{22}\pm\sqrt{\left(C_{11}+C_{22}\right)^2
-4\left(C_{11}C_{22}-C_{12}C_{12}\right)}\right], \nonumber\\
\lambda_3 = C_{33}. \label{lambda}
\end{gather}
Equivalent expressions are found for $\lambda^S_{i}$ by substituting $C^S_{ij}$ instead of $C_{ij}$ into \eq{lambda}.

In Fig.~\ref{f1} we compare the eigenvalues in \eq{lambda} with the corresponding relativistic values $\lambda^S_i$ for a parabolic trajectory with $h=5\,\rs\,\cc$ (which represents a $\beta \simeq 1$ encounter between a solar-type star and a $10^6\,M_{\odot}$ black hole). As a reference to illustrate the importance of relativistic effects, we have also plotted the corresponding Newtonian values \citep{brassart08}
\begin{equation}
\lambda_1^N = \frac{2\,\G M_{\rm bh}}{r^3},\qquad \lambda_2^N = \lambda_3^N = -\frac{\G M_{\rm bh}}{r^3}.
\end{equation}
In this figure and for any other parabolic encounter, the eigenvalue $\lambda_3$ coincides exactly with the corresponding relativistic result. More specifically, for a parabolic trajectory with angular momentum $h$, it is found that
\begin{equation}
\lambda_3 = \lambda^S_3 = -\frac{\G M_{\rm bh}}{r^3}
\left(1-\frac{\rs}{r}\right)^2\left(1+\frac{3\,h^2}{r^2\cc^2}\right).
\end{equation}
This means that tidal compression along the vertical direction is reproduced exactly by the \citetalias{tejeda13} potential for parabolic encounters.
On the other hand, it is also apparent from Fig.~\ref{f1} that the eigenvalues $\lambda_1$ and $\lambda_2$ provide a good approximation to the exact relativistic values.

\begin{figure}
\begin{center}
\includegraphics[width=\linewidth]{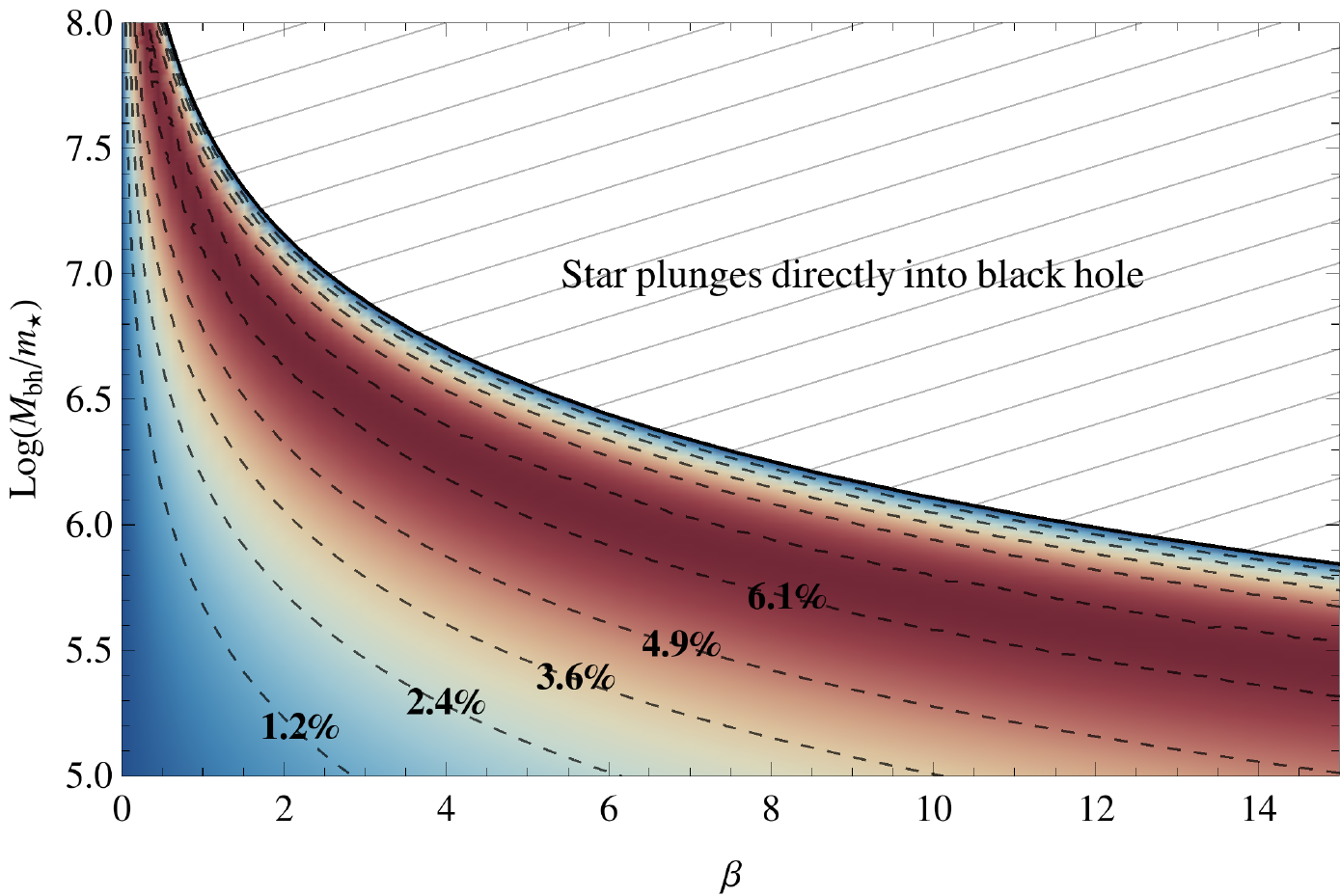}
\end{center}
\caption{Contours of the maximum relative error with which the eigenvalue of the tidal tensor $\lambda_1$ \eqp{lambda} reproduces the exact relativistic value $\lambda^{S}_1$ as a function of the impact parameter $\beta$ and the mass ratio $q=M_{\rm bh}/m_\star$.}
\label{f2}
\end{figure}
In a practical simulation, the maximum departure of the tidal field corresponding to the \citetalias{tejeda13} potential from the relativistic result is reached at the periapsis of the incoming trajectory. In Fig.~\ref{f2} we show the contours of the maximum relative error for $\lambda_1$ (i.e. $(\lambda^S_1-\lambda_1)/\lambda^S_1$) as a function of the impact parameter $\beta$ and the mass ratio $q=M_\mathrm{bh}/m_\star$. From this figure we see that expansion due to the tidal tensor is reproduced by the \citetalias{tejeda13} potential with a precision better than $6.4\,\%$. Finally, the relative error for $\lambda_2$ at periapsis was found to be consistently zero within machine precision ($\approx10^{-15}$) for the parameter values shown in Fig.~\ref{f2}. This appears to contradict the fact that $\lambda^{TR}_2$ deviates from $\lambda^{S}_2$ at $r\sim 2\rs$ in Fig.~\ref{f1}, but note that the maximum error occurs at pericentre, which in Fig.~\ref{f1} is at $r\sim 24\rs$, and that the curves shown there will change with $\beta$.

\section{The Roche potential for non-circular orbits}
\label{sec:app2}

The Roche lobe is defined in the context of the so-called restricted
three-body problem, where a test particle moves in the potential of two
orbiting masses. It is bounded by the
critical equipotential surface of the effective potential 
$\Phi_{\rm eff}(\mathbfit{r})$, incorporating inertial forces in the coordinate
frame that is comoving and corotating with the star \citep{sepinsky07}.
In such a non-inertial frame, the expression for the effective potential reads:
\begin{equation}
\Phi_{\rm eff}(\mathbfit{r}) 
  = \Phi_*(\mathbfit{r}) 
  + \Phi_{\rm bh}(\mathbfit{r}) 
  - (\mathbfit{r}-\mathbfit{r}_*) \cdot {\mathbf{\nabla}_*}\Phi_{\rm bh}(\mathbfit{r}_*) 
  - \frac{1}{2}|\mathbf{\omega}(\mathbfit{r})\times(\mathbfit{r}-\mathbfit{r}_*)|^2.
\end{equation}
Here, $\mathbfit{r}_*$ is the current instantaneous position of the centre of mass
(CoM) of the star, $\mathbf{\nabla}_*$ is the gradient with respect to $\mathbfit{r}_*$,
$\mathbf{\omega}(\mathbfit{r})$ is the angular velocity at the position $\mathbfit{r}$. 
The first two terms in this expression, $\Phi_*(\mathbfit{r})$ and 
$\Phi_{\rm bh}(\mathbfit{r})$, represent the gravitational potentials of the star
and the black hole, respectively; in our simulations, the former is computed with the tree, 
while the latter can be either $\Phi_{\rm N}$ or $\Phi_{\rm TR}$.
The third term appears because our reference frame is comoving with the star.
It produces uniform acceleration, equal and opposite to the one of the CoM of
the star.
The last term is the centrifugal potential due to stellar rotation.

If we use point particle Newtonian potentials for the black hole and the star,
the effective potential becomes:
\begin{equation}
\Phi_{\rm eff}(\mathbfit{r}) 
  = -\frac{G m_*}{r} 
  -  \frac{G M_{\rm bh}}{r}
  - \frac{G M_{\rm bh}}{r_*^3}\, \mathbfit{r}_* \cdot (\mathbfit{r}-\mathbfit{r}_*) 
  - \frac{1}{2}|\mathbf{\omega}(\mathbfit{r})\times(\mathbfit{r}-\mathbfit{r}_*)|^2.
\end{equation}

The choice of angular velocity $\mathbf{\omega}$ in the effective potential depends on
how the star is rotating and is not very clearly defined in the
post-disruption phase.
Therefore, in Fig.~\ref{fig:Roche}, for the contours of  $\Phi_{\rm eff}(\mathbfit r)$
(dashed black lines) we adopt an average angular velocity 
$\langle\mathbf{\omega}\rangle = \frac{1}{N_{\rm part}} \sum_a\mathbf{\omega}_a$ (where $a$ is the particle index), 
while for the colours of individual particles we use the values of the 
individual angular velocities $\omega_a$.

\label{lastpage}
\end{document}